\documentclass[%
 aip,
 rsi,
 amsmath,amssymb,
 reprint,%
 floatfix
]{revtex4-1}

\usepackage[utf8]{inputenc}
\usepackage{setspace}
\usepackage{graphicx}
\usepackage{wrapfig}
\usepackage{booktabs}
\usepackage{makecell}
\usepackage{dcolumn}
\usepackage{amsmath}
\usepackage{etoolbox}

\usepackage[mathlines]{lineno}

\newcolumntype{d}[1]{D{.}{.}{#1} }

\draft 

\begin{document}

\title{Optimization of Ray-tracing Simulations to Confirm Performance of the GP-SANS Instrument at the High-Flux Isotope Reactor}

\author{James M. Rogers}
 \email{jroger87@vols.utk.edu}
 \affiliation{The University of Tennessee, Knoxville}

\author{Matthew J. Frost}
 \email{frostmj@ornl.gov}
 \affiliation{ORNL}

\author{Lisa M. DeBeer-Schmitt}
 \email{debeerschmlm@ornl.gov}
 \affiliation{ORNL}

\date{\today}

\begin{abstract}  
The CG-2 beamline at the High Flux Isotope Reactor (HFIR) exhibits a notable discrepancy between observed count rates and the count rates we would expect based on a Monte-Carlo neutron ray-trace simulation.
These simulations consistently predict count rates approximately five times greater than those observed in four separate experimental runs involving different instrument configurations.
This discrepancy suggests that certain factors are causing losses in measurements that are not adequately accounted for in the simulation, in particular guide reflectivity or misalignment.

To investigate these discrepancies, a high-dimensional simulation parameter approach is applied in order to understand the losses.
Region of Interest (ROI) groups along the instrument are assigned to different surfaces of the guide components within the simulation.
This allows the parameters of those guide components to be varied as a group to minimize the complexity of the search space.
The result is an optimization of simulation parameters using an iterative scheme that aims to minimize the difference between experimentally measured count rates and simulated count rates across all tested collimator combinations.

This proposed methodology holds the potential to reveal previously unrecognized sources of intensity loss in the CG-2 beamline at HFIR and improve the accuracy of simulations, leading to enhanced understanding and performance of the beamline for various scientific applications.
\end{abstract}

\maketitle
\noindent {\em \small Notice: This manuscript has been authored by UT-Battelle, LLC, under contract DE-AC05-00OR22725 with the US Department of Energy (DOE). The US government retains and the publisher, by accepting the article for publication, acknowledges that the US government retains a nonexclusive, paid-up, irrevocable, worldwide license to publish or reproduce the published form of this manuscript, or allow others to do so, for US government purposes. DOE will provide public access to these results of federally sponsored research in accordance with the DOE Public Access Plan (https://www.energy.gov/doe-public-access-plan).}


\section{Introduction \label{sec:introduction} }

The General-Purpose Small Angle Neutron Scattering (GP-SANS) Instrument, situated at the CG-2 beam port within the Cold Guide Hall of the High Flux Isotope Reactor (HFIR) which is operated by  Oak Ridge National Laboratory, has delivered noteworthy experimental outcomes in the fields of chemistry, engineering, superconductivity, magnetic materials and biology spanning almost two decades.
Although widely acknowledged for its world-class capabilities, there is discrepancy between the actual instrument performance as compared to the study documented by Moon et al. \cite{moonHB4Brightness}.
Most notably, the intensity is different than expected when measured using low efficiency counters.

In May 2023, a comprehensive set of intensity measurements were carried out with a low-efficiency gas counting detector in the sample position using a variety of upstream instrument configurations, with the primary objective of quantifying the overall beam characteristics.
Additionally, over the past five years a simulation of the instrument's current design has been developed.
This simulation serves a dual purpose: it aids in the conceptualization of potential upgrades to the instrument and provides a valuable tool for users.

The overarching aim  of this work is to integrate the data obtained from recent beam characterization measurements and the current state of the model simulation, with the intention of creating a precise virtual representation of the instrument. This comprehensive approach seeks to not only provide an accurate depiction of the instrument but also identify the specific components accountable for the observed discrepancy. 

\subsection{CG-2 Beamline}
The CG-2 beamline extends off of Horizontal Beamtube 4 (HB-4) at HFIR, which delivers neutron flux from the beryllium reflector surrounding the reactor core through a 20 K liquid hydrogen moderator vessel designed to maximize 4 - 12 Å neutron flux. 
Once neutrons pass through the beamtube, an internal collimator separates the flux into several channels corresponding to the different instruments in the Cold Guide Hall, with CG-2 using Channel 2. 
Neutrons must also pass through a rotary shutter fabricated using carbon steel and high density concrete.
This is provides shielding when not in use, and an external collimator system before passing on to the CG-2 guide itself.

From there, neutrons pass through a 4.0 meter long 4.0 x 4.0 cm guide with polished nickel surfaces which delivers them to an ``optical filter'' angled at 1° relative to the first guide.
After passing 2.292 m through the optical filter, neutrons go through a second guide that is 18.9 m long angled another 1° relative to the optical filter.
An optional helical disk type velocity selector then filters out undesired wavelengths before neutrons pass through eight removable guide sections each 2.0 meters long which will be referred to going forward as instrument collimators. 
After traveling 18.0 meters past the exit of the velocity selector, an exit slit limits the beam diameter incident on the sample position.
The GP-SANS detector itself is a one meter wide array of 192 one meter long 3-Helium Linear Position Sensitive Detectors (LPSD) \cite{berry_characterization_2012} on rails in a vacuum chamber, allowing users to vary the sample to detector distance between 1 - 19 m. 

The HB-4 source and beam tube were characterized during commissioning\cite{robertson_measured_2008} and contain no reflection components.
Because of this fact, only CG-2 beamline components were were the subject of optimization.

The reflectivity parameters for all guides within the instrument are set at $m=1$, except for the optical filter. The reflecting side of the optical filter is specified with $m=3$, while its opposite side acts as an absorber with $m=0$. The top and bottom sections of the optical filter maintain a reflectivity parameter of $m=1$.
A more detailed discussion about reflectivity parameters and their impact on instrument performance will be had in sections \ref{sec:measurement} and \ref{sec:simulation}.
Comparing a revised analysis \cite{FrostHB4Brightness} of a measurement of the cold source in 2007 \cite{robertson_measured_2008} to the cold source performance as described by an internal memorandum in December 2000 \cite{moonHB4Brightness} shows a 25\% discrepancy in the anticipated cold source brightness.
This has already been accounted for in the simulation as the revised triple Maxwell-Boltzman brightness and temperature parameters have been used in the source component.


\section{Measurement \label{sec:measurement} }

\begin{figure}
  \centering
  \includegraphics[width=\linewidth]{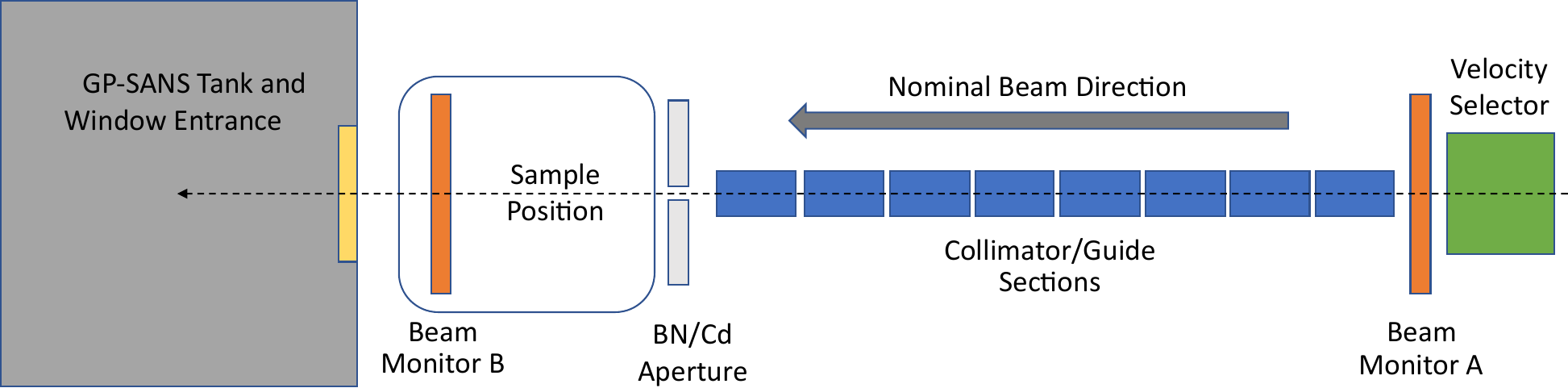}
  \caption{A schematic describing the of Layout at CG-2 during the measurement.}
  \label{fig:measurementSchematic}
\end{figure}

In May 2023, a series of white beam measurements were conducted at the GP-SANS sample position (Beam Monitor B, FIG \ref{fig:measurementSchematic}), exploring various guide and aperture configurations.
These measurements involved the removal of the velocity selector typically utilized in standard operations.
To define the beam at the sample position, a set of circular apertures with diameters of 6, 10, and 20 mm were used in combination with instrument collimator settings of 8, 7, 4, and 0 sections. 
Notably, there exists a 4 cm square aperture immediately upstream of the instrument collimator sections.
The measurement at the sample position was taken with a neutron sensitive gas proportional gas counter fabricated by ORDELA \cite{ORDELA}, model 4500 S/N 002 with listed detection efficiency of $10^{-5}$ counts/neutron at 1.8~\AA.

\begin{figure}
  \centering
  \includegraphics[width=\linewidth]{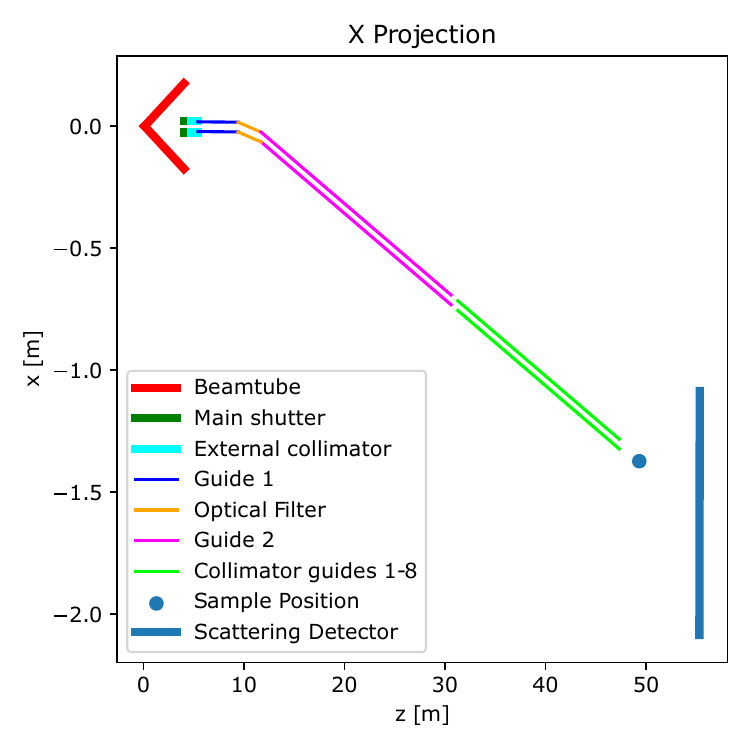}
  \caption{A visualization of the HB-4 beam tube and the CG-2 instrument. After exiting the HB-4 beam tube, the neutrons are guided towards the target by the Optical Filter component, then towards the Velocity Selector.
  Finally, it passes through a series of 0-8 removable beam guides before reaching the sample position. This can be seen with better detail in Figure \ref{fig:measurementSchematic}.}
  \label{fig:HB-4_and_CG-2}
\end{figure}

In addition, data from Beam Monitor A (Fig \ref{fig:measurementSchematic}) was extracted from experiments performed in December of 2022. 
This monitor is an ORDELA of the same model as the one at the sample position but with a measured neutron detection efficiency $1.09\times10^{-5}$ at 1.8~\AA.
This monitor is located immediately downstream of the velocity selector and the attenuator package(not shown) and always available during normal user operation as denoted in Fig \ref{fig:measurementSchematic}.
This data is used to understand the instrument from the source  to the velocity selector exit, which will also include any transmission and bandwidth effects introduced by the velocity selector.
All of this data was reduced and analyzed for comparison to the simulation as well as prior analytical estimations on performance\cite{moonCGSystemPerformance}. 

\if False { 
\begin{table}[h!]
  \begin{center}
    \caption{Velocity Selector Measurements}
    \label{tab:table1}
    \begin{tabular}{c|d{1}|d{1}}
      \toprule
      \textbf{VS Angular Velocity [RPM]} & \textbf{Measured Count Rate [c/s]}\\
      \midrule
        6000 & 22316.7701714883 \\
        5800 & 23173.3034424436 \\
        5600 & 23982.8401443759 \\
        5400 & 24809.8731865166 \\
        5200 & 25451.618593222 \\
        5000 & 25807.6854776222 \\
        4800 & 25786.844542551 \\
        4600 & 25360.1736975112 \\
        4400 & 24663.5121023016 \\
        4200 & 23754.901286091 \\
        4000 & 22733.4856536928 \\
        4000 & 22798.976157206 \\
        3900 & 22255.7167456345 \\
        3800 & 21661.7019638333 \\
        3700 & 21031.581320021 \\
        3600 & 20422.2517819795 \\
        3500 & 19688.5080136395 \\
        3400 & 18926.521024539 \\
        3300 & 18122.0092098631 \\
        3200 & 17209.0493840665 \\
        3100 & 16292.8011085474 \\
        3000 & 15354.6262029369 \\
        2900 & 14326.8277970337 \\
        2800 & 13328.0202040933 \\
        2700 & 12316.0273802401 \\
        2600 & 11351.204070155 \\
        2500 & 10416.40443605 \\
        2400 & 9501.46983577896 \\
        2300 & 8610.28561273619 \\
        2200 & 7731.35361902613 \\
        2100 & 6917.77843418989 \\
        2000 & 6118.89698345556 \\
        1900 & 5355.6799526462 \\
        1800 & 4623.08411691631 \\
        1700 & 3913.23234916212 \\
        1600 & 3213.23103387066 \\
      \bottomrule
    \end{tabular}
  \end{center}
  \caption{Measured count rate (in counts per second) for Velocity Selector Rotation Speeds. Statistical error is no more than 0.5\% for any of the measurements.\label{tab:vsCountrates}}
\end{table}
} \fi

\begin{figure}
  \includegraphics[width=\linewidth]{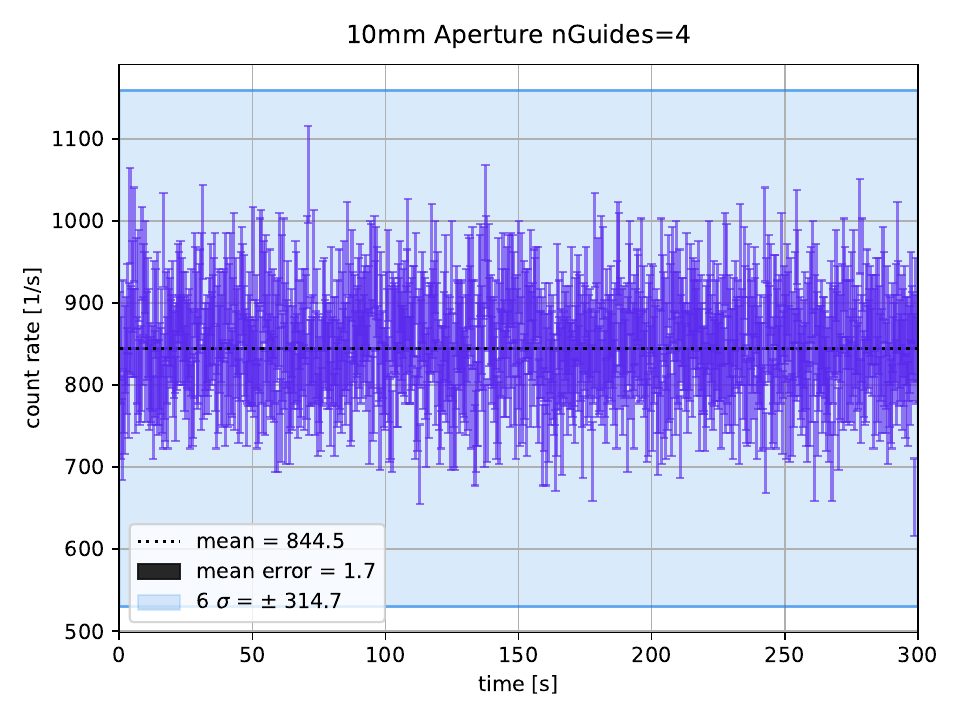}
  \caption{Measured count rate at the CG-2 sample position. After rebinning to 0.3 s per bin, data points beyond six standard deviations were assumed to be outliers from normal operation, and were excluded in the determination of the mean count rate.\label{fig:measuredCountRates}}
\end{figure}

\if {false}
\begin{figure}
  \includegraphics[width=\linewidth]{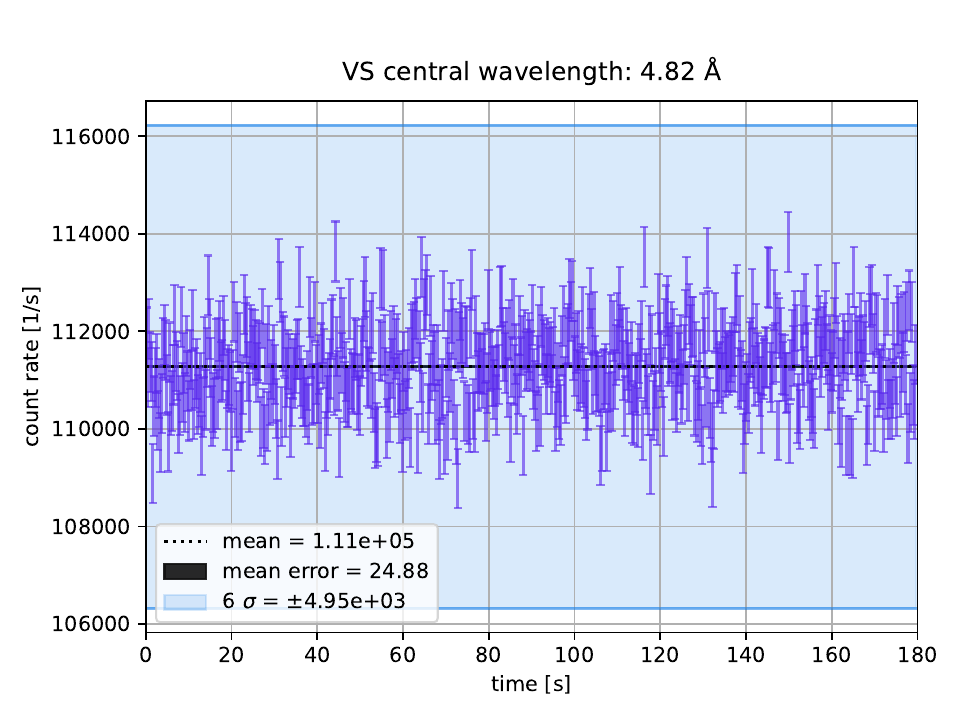}
  \caption{Measured count rate at Beam Monitor A, immediately after the velocity selector. After rebinning to 0.3 s per bin, data points beyond six standard deviations were assumed to be outliers from normal operation, and were excluded in the determination of the mean count rate.\label{fig:measuredCountRates}}
\end{figure}
\fi

All monitor data was recorded in the NeXuS-HDF5 file format \cite{PETERSON201524}, which was then extracted using the h5py Python module \cite{collette_python_hdf5_2014}.
The data sets are binned in time to check for consistent source operation, and outliers past six sigma were removed to determine the average count rate without intensity variation, as seen in Figure \ref{fig:measuredCountRates}.
The average count rate for each run is calculated as a weighted mean across the included data points, and count rates corresponding to the various aperture and collimator configurations are seen in Table \ref{tab:whitebeamCountrates}.
\begin{table}[tb]
  \begin{center}
    \begin{tabular}{c||c|c|c|c}
      \toprule
      \textbf{Aperture\textbackslash Collimators} & \textbf{0} & \textbf{4 } & \textbf{7} & \textbf{8}\\
      \hline
      6 mm & 91.0 & 292 & 1460 & 1780 \\
      10 mm & 264 & 845 & 4090 & 5130 \\
      20 mm & 1030 & 3230 & 14700 & 19400 \\
      \hline
    \end{tabular}
  \end{center}
  \caption{Measured count rate by Beam Monitor B (in counts per second) for different collimator and aperture configurations. Statistical error is no more than 0.5\% for any of the measurements.}
  \label{tab:whitebeamCountrates}
\end{table}
For white beam measurements (Monitor B, at the sample position), count rates are proportional to the exit aperture diameter squared, and increasing the number of instrument collimator sections permits more neutrons to be transported to the sample position, resulting in a higher count rate.
The length of time taken for each run permitted the relative measured error of the intensity to be less than 0.5\% based on counting statistics.
\section{Simulation \label{sec:simulation} }

\subsection{McStas for Neutron Raytracing}

McStas \cite{willendrup_mcstas_2020} is a widely adopted Monte Carlo neutron ray-tracing simulation tool utilized globally for designing and analyzing neutron instruments at scattering facilities. Its flexibility, customization, and open-source nature have made it an industry standard. It offers robust tools for users to adjust neutron sources, beam optics, samples, and detectors, enabling refinement and optimization of instrument designs.
Originally developed at Risø-ILL, McStas now hosts a wide range of both developer written and user contributed components found in ``.comp'' files \cite{mcstas_component_manual}. 
These can be easily included and utilized in ``.instr'' files, which contain the main instructions for assembly of  instruments.

McStas excels in neutron optics simulation, and the ability to modify parameters and the underlying source code makes it easy to achieve highly accurate predictions on a wide range of observable quantities as well as some quantities that cannot be directly observed, such as specific spin vector or trajectory components. 
For example, McStas simulations have detectors that are 100\% efficient, and neutrons can be measured without affecting their state.
This contrasts with real-world constraints imposed by the Heisenberg uncertainty principle, limiting the precision of simultaneously measuring certain properties such as position and momentum.
Additionally, rather than utilizing limited resource of beam time to explore component or sample parameters, computational time can be employed to offer estimations of desired quantities. This approach aids in making informed decisions regarding necessary upgrades or refurbishments for components.
However, while McStas is capable of accurate simulation in many aspects of neutron optics, it does not have the ability to model quantum phenomena, thus limiting it mainly to instrument design rather then total experiment replication.
For this purpose, it may be necessary to use additional modeling tools such as MCViNE \cite{lin_mcvine_2019}, MCNP \cite{kulesza_mcnp_2022}, or Geant4 \cite{allison_geant4_2006} to understand precise nuclear or material phenomena.

McStas is still best suited for these investigations given the ability to complete the necessary calculations faster and high community confidence in understanding the instrument components have on the neutron beam character.
This is seen from prior experience in instrument development and application.

\subsection{McStas Implementation of CG-2 Beamline}

\if{
Originally, the CG-2 beamline simulation was written in McStas version 2, but the only changes needed to update this to version 3.2 were to modify a few deprecated variable names and switch out the obsolete horizontal divergence ``Hdiv monitor'' component for the new ``Div1D monitor'' \cite{cg2_ornl_repo_2023}.
While those changes are small and have no real impact on the over capability or result, the decision to keep the design up to date with the latest version of the toolkit is appropriate given that the team hopes to have this version of the simulation available to a wide range of users.
}
\fi

This section will be a short discussion of the individual components used in the simulation of the CG-2 instrument at HFIR.  These components are shown in Fig. \ref{fig:HB-4_and_CG-2}.  For this instrument build, CG-2 is made up of Guide 1 which transports the neutrons from the external collimator in the HB-4 beam tunnel, then is immediately followed by the optical filter.  The main neutron transport to the instrument hall is done through Guide 2.  Between Guide 2 and the Collimation section is the instrument velocity selector.  The neutrons are then transported to the sample position and then on to either a beam monitor B as shown in Fig. \ref{fig:measurementSchematic} or the instrument detector.

The Guide component is used extensively in the simulation and it is part of the original component library written by Krisian Nielsen (Risø), giving confidence that it has been extensively tested and vetted by all McStas users.
In contrast, the optical filter component is an ORNL developed component based off of the Guide component, but modified by J. Lee Robertson (retired, formerly ORNL) to match performance of the optical filter installed at CG-2.  
It is essentially the regular Guide component but modified so that each m value on the interior faces is independently variable.
This allows the left vertical surface to be modeled as a supermirror (per Equation \ref{eq:McStas_StdReflecFunc}) with normalized $Q_{Ni}$ critical edge m=3, while the top and bottom horizontal surfaces can be m=1, and the right vertical surface is an absorber with m=0. 

The ``Helical multidisk'' type velocity selector in use at CG-2 is unique in that that it is comprised of a series of slotted gadolinium coated disks rather than the standard ``helical lamella'' type \cite{friedrich_high-performance_1989}. 
The main benefit of this design is that the selector is much lighter than the lamella based ones, allowing the selector to spin at a much faster rate to provide shorter wavelengths \cite{hammouda_multidisk_1992}. 
The McStas component used in the simulation to reproduce the disk type velocity selector was also developed at ORNL by J. Lee Robertson in 2018.
It works by allowing the neutron to pass through the first disk with a random disk phase angle, then propagating the ray through the velocity selector and determining if the neutron is absorbed by the last disk using propagation time and the phase angle of the first disk. 
The probability of the neutron passing through the first disk is multiplied by the weight of unabsorbed neutrons, efficiently using neutron rays while preserving statistical properties of the velocity selector.  

The final major component for the CG-2 beam line before getting to the sample area and detectors are the collimator sections, made up of 8 Guide components that can either be put into the beam for more flux on sample or removed to give higher resolution as needed.  These current guide pieces have m=1 coatings and were installed in 2019 at the beam line as an upgrade. They are also non-magnetic for the future upgrade of polarization on the beam line.

\begin{figure}[bt]
  \centering
  \includegraphics[width=\linewidth]{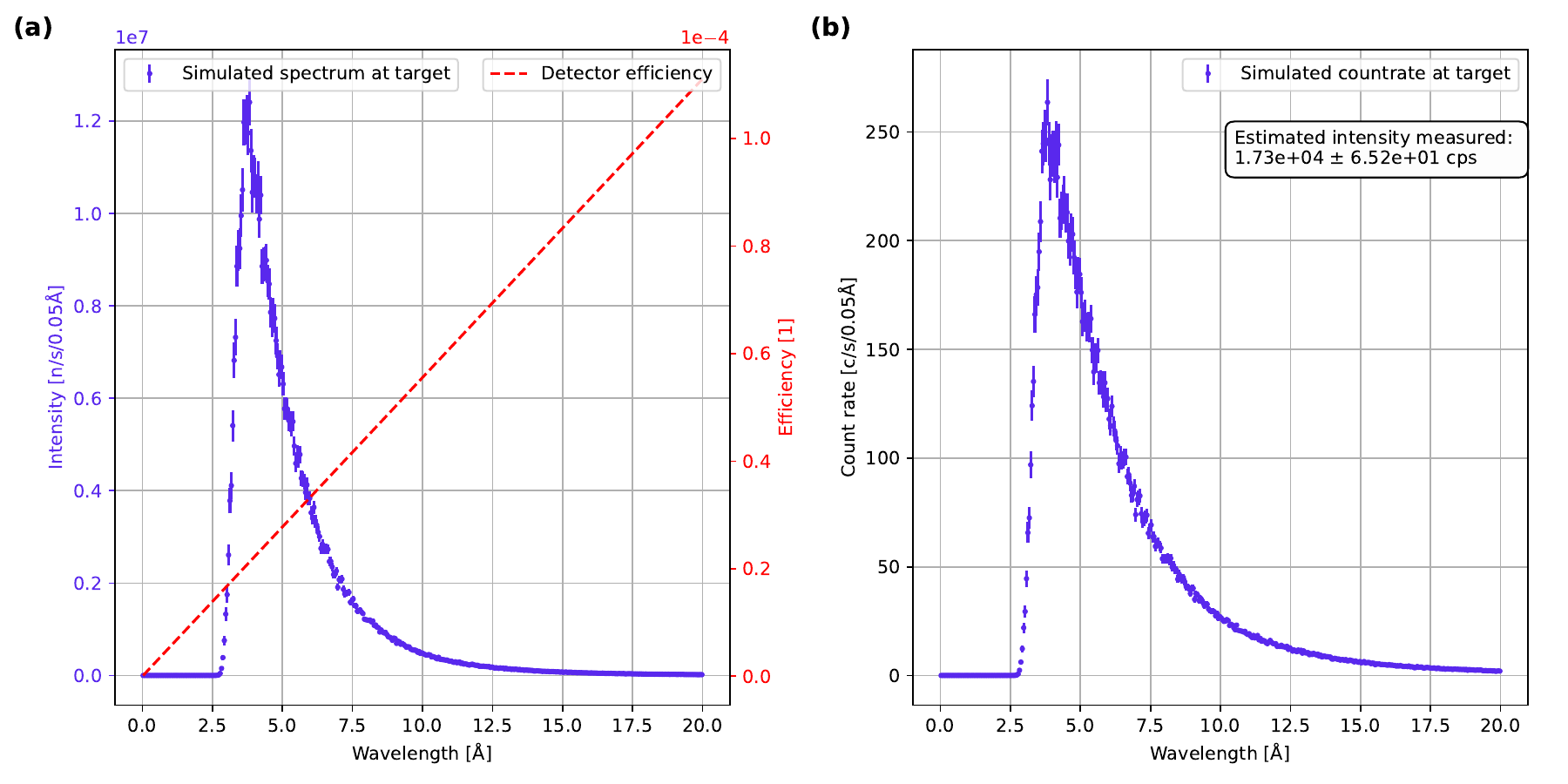}
  \caption{In (a), simulated intensity spectrum collected at sample position by beam monitor B (blue) is compared to the corresponding detector efficiency in the linear regime (red). For beam monitor B, this detector efficiency is $1.00 \times10^{-5}$ counts per 1.8\AA~neutron. 
  (b) shows how simulated countrate spectrum was derived from the product of these two curves. \label{fig:simulated_intensity_spectrum}}
\end{figure}

The simulated intensity spectrum at the sample position was scaled by a known detector efficiency factor of $1.00 \times10^{-5}$ counts per 1.8\AA~neutron shown in red in Figure \ref{fig:simulated_intensity_spectrum}.
This scaled countrate was then compared to the measurement taken using ``Beam Monitor B'' denoted in Figure \ref{fig:HB-4_and_CG-2}.
The detector immediately downstream of the velocity selector was slightly different and has a neutron detection efficiency of $1.09 \times10^{-5}$ per 1.8\AA~neutron.
This is designated as ``Beam Monitor A''  also in Figure \ref{fig:measurementSchematic}.
The scaled spectrum corresponds to the spectrum countrate [c/s] of the real detector, and the estimated total count rate by simulation is taken as the sum of countrate across that scaled spectrum.

\subsection{Simulation Discrepancy}

For each of the different instrument configurations, including the white beam measurement and each velocity selector speed, a simulation with matching parameters and data collection points was completed. 
Simulation count rate was estimated by scaling the monitor spectra for these runs by their corresponding detector efficiency and integrating the resulting spectrum, as seen in Figure \ref{fig:simulated_intensity_spectrum}. 
These estimations show a nominal factor of 5.1 greater count rate as compared to white beam measurements taken at the sample position, and a factor 1.4 greater than monochromatic measurements taken at the exit of the velocity selector, suggesting that the simulation needs to be adjusted to more accurately represent the beam line.

\begin{figure*}
  \centering
  \includegraphics[width=\linewidth]{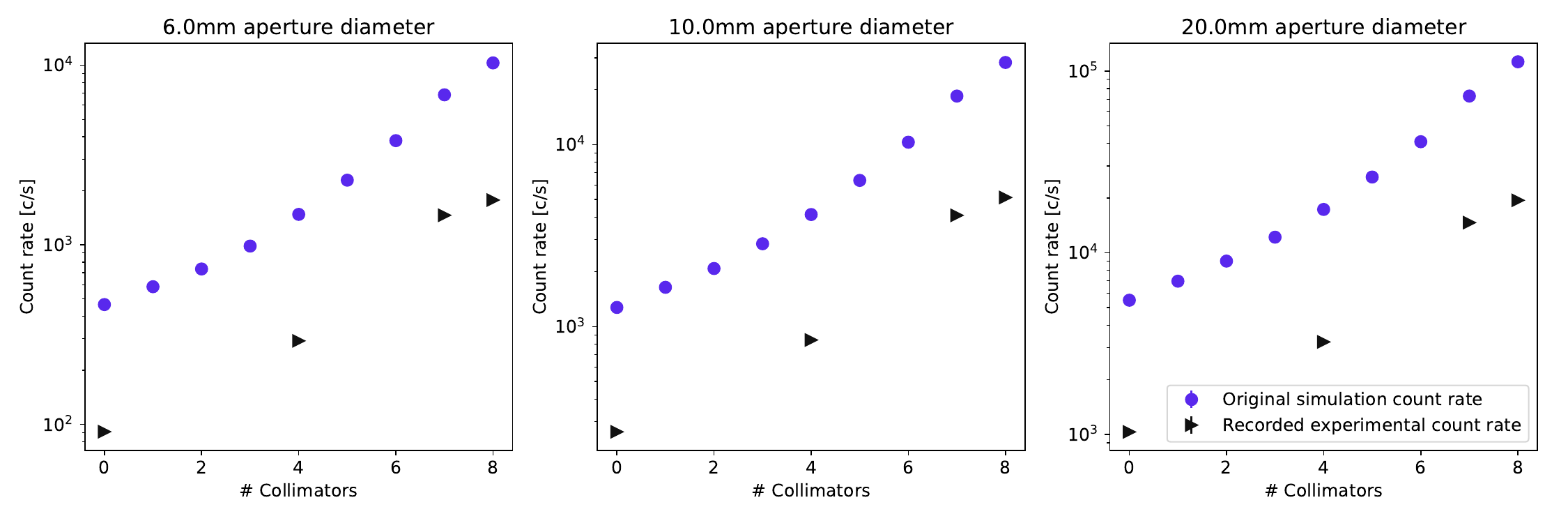}
  \caption{A comparison of experimental measurements to simulation countrates. The measurements taken at the Sample Position (black) are on average a factor of 5.2 lower than the simulation countrates (blue) predict. See also Table \ref{tab:whiteBeamDisc}.}
  \label{fig:White_beam_measured_countrates}
\end{figure*}

\begin{table*}
  \begin{center}
    \caption{White Beam Measurement Discrepancy. The table shows the difference between the measured rate and the simulated rate at the sample position.}
    \label{tab:whiteBeamDisc}
    \begin{tabular}{c|c|d{1}|d{1}|d{1}}
      \toprule
      \textbf{\# Collimators} & \makecell{\textbf{Exit Aperture}} & \makecell{\textbf{Measured}} & \makecell{\textbf{Simulated}} &      \makecell{\textbf{Ratio}}\\
      \textbf{} & \makecell{\textbf{Diameter [mm]}} & \makecell{\textbf{Rate [c/s]}} & \makecell{\textbf{Rate [c/s]}} & \makecell{\textbf{}}\\
      \hline
      0 & 6 & 91.0 & 464.8 & 5.1\\
      4 & 6 & 291.7 & 1477.6 & 5.1\\
      7 & 6 & 1461.7 & 6848.2 & 4.8\\
      8 & 6 & 1775.1 & 10311.3 & 5.8\\
      \hline
      0 & 10 & 264.3 & 1274.9 & 4.8\\
      4 & 10 & 844.5 & 4130.9 & 4.9\\
      7 & 10 & 4087.6 & 18485.1 & 4.5\\
      8 & 10 & 5126.1 & 28281.4 & 5.5\\
      \hline
      0 & 20 & 1032.9 & 5480.6 & 5.3\\
      4 & 20 & 3230.0 & 17311.4 & 5.4\\
      7 & 20 & 14652.6 & 72827.8 & 5.0\\
      8 & 20 & 19430.6 & 112422.8 & 5.8\\
      \hline
    \end{tabular}
  \end{center}
\end{table*}

Figure \ref{fig:White_beam_measured_countrates} shows a graphical comparison of this discrepancy for white beam measurements. The experimental data is represented by the black triangles and the simulated data is by the blue for a few examples of the instrument configurations.  Table \ref{tab:whiteBeamDisc} shows all the instrument configurations white beam data sets and the simulations.  
\begin{figure}[bt]
  \centering
  \includegraphics[width=\linewidth]{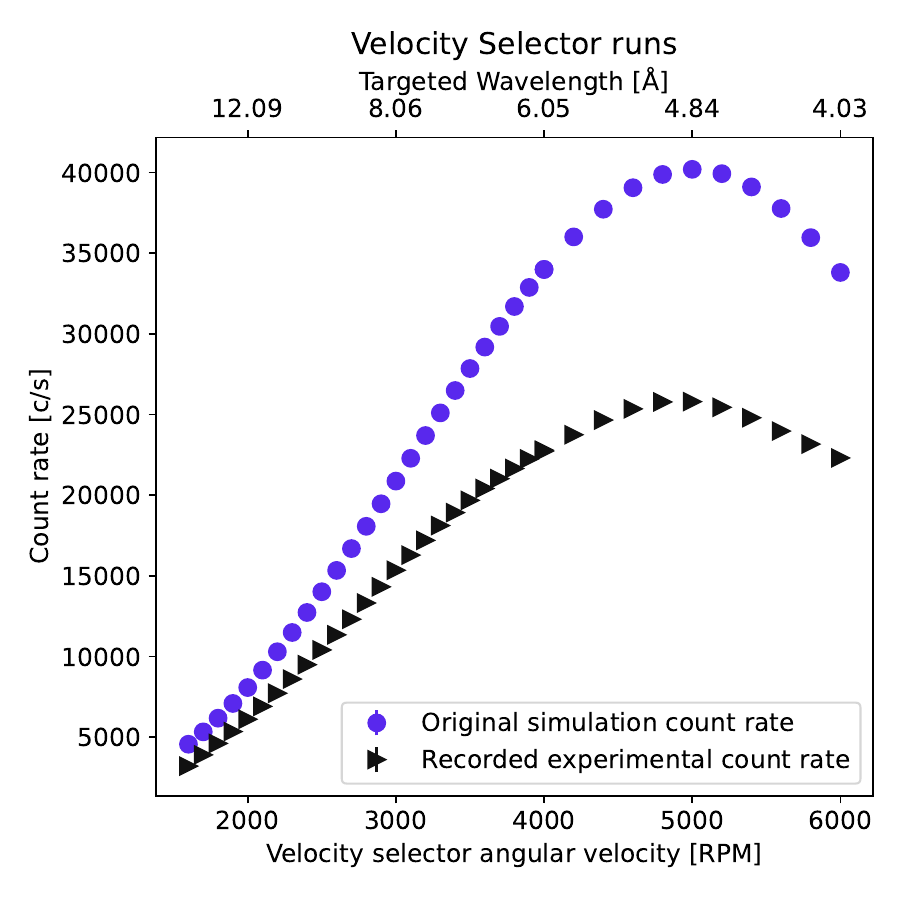}
  \caption{A comparison of simulated and actual count rates at the exit of the velocity selector under different operating speeds. An average factor of 1.43 difference between simulation and experimental intensity measurement immediately downstream of the Velocity Selector.}
  \label{fig:Velocity_Selector_measured_countrates}
\end{figure}

\if False { 
\begin{table}[h!]
  \begin{center}
    \caption{Velocity Selector Measurement Discrepancy}
    \label{tab:table1}
    \begin{tabular}{c|d{1}|d{1}}
      \toprule
      \textbf{VS Angular Velocity [RPM]} & \textbf{Measured Count Rate [c/s]} & \textbf{Estimated Count Rate [c/s]}\\
      \midrule
        6000 & 22316.7701714883 & 33814.0464541097 \\
        5800 & 23173.3034424436 & 35973.0973739617 \\
        5600 & 23982.8401443759 & 37778.302483393 \\
        5400 & 24809.8731865166 & 39113.9365468413 \\
        5200 & 25451.618593222 & 39937.0832329818 \\
        5000 & 25807.6854776222 & 40204.2741037116 \\
        4800 & 25786.844542551 & 39887.1807972261 \\
        4600 & 25360.1736975112 & 39065.8704724702 \\
        4400 & 24663.5121023016 & 37732.6253167956 \\
        4200 & 23754.901286091 & 36017.4270074276 \\
        4000 & 22733.4856536928 & 33999.9982583725 \\
        4000 & 22798.976157206 & 33999.9982583725 \\
        3900 & 22255.7167456345 & 32886.2310534581 \\
        3800 & 21661.7019638333 & 31702.5721874034 \\
        3700 & 21031.581320021 & 30470.6024674834 \\
        3600 & 20422.2517819795 & 29187.9229784844 \\
        3500 & 19688.5080136395 & 27859.4791548013 \\
        3400 & 18926.521024539 & 26495.731263292 \\
        3300 & 18122.0092098631 & 25108.9235191196 \\
        3200 & 17209.0493840665 & 23705.1097159722 \\
        3100 & 16292.8011085474 & 22293.9104362884 \\
        3000 & 15354.6262029369 & 20882.1358165992 \\
        2900 & 14326.8277970337 & 19475.0625462354 \\
        2800 & 13328.0202040933 & 18078.6715073231 \\
        2700 & 12316.0273802401 & 16699.6950001111 \\
        2600 & 11351.204070155 & 15345.6242119309 \\
        2500 & 10416.40443605 & 14023.5953477965 \\
        2400 & 9501.46983577896 & 12739.4808435019 \\
        2300 & 8610.28561273619 & 11498.1356355343 \\
        2200 & 7731.35361902613 & 10304.3838412796 \\
        2100 & 6917.77843418989 & 9163.94398677789 \\
        2000 & 6118.89698345556 & 8084.85949611912 \\
        1900 & 5355.6799526462 & 7100.24132388409 \\
        1800 & 4623.08411691631 & 6186.94902571876 \\
        1700 & 3913.23234916212 & 5337.31881767818 \\
        1600 & 3213.23103387066 & 4571.17572496546 \\
      \bottomrule
    \end{tabular}
  \end{center}
\end{table}
} \fi

Figure \ref{fig:Velocity_Selector_measured_countrates} shows this discrepancy for measurements taken with the velocity selector, with an average factor of 1.43 difference. 
The discrepancies are hypothesized to be due to either misalignment of the guide components or possibly degradation of the m-coating on the upstream guides from aging.
\section{Parameter Optimization \label{sec:optimization} }

To test the validity that guide degradation or misalignment could be the primary causes of the intensity discrepancy, the first-order reflectivity parameters ($R_0$, $m$, $\alpha$) of the surfaces in the CG-2 simulation were varied and components were given misalignment parameters for all six degrees of freedom which will be discussed in more detail in later sections.
The tolerances of these parameters were investigated, and then a full scale optimization was run to minimize the intensity discrepancy between the simulation and the measurements.

\subsection{Variable Parameters}
\subsubsection{Reflectivity Parameters}
Each reflecting surface in the instrument has an associated reflectivity curve describing the probability of a neutron with incident wave-vector $q$ being reflected. 
The equation for this reflectivity curve is given by \cite{willendrup_mcstas_2020}

\begin{eqnarray}
R(q) = \frac{R_0}{2}\left(1 - \tanh\left(\frac{q - m \cdot Q_c}{w}\right)\right) \times \left(1 - \alpha \cdot (q - Q_c)\right)
    \label{eq:McStas_StdReflecFunc}
\end{eqnarray} 

\begin{figure} 
  \centering
  \includegraphics[width=\linewidth]{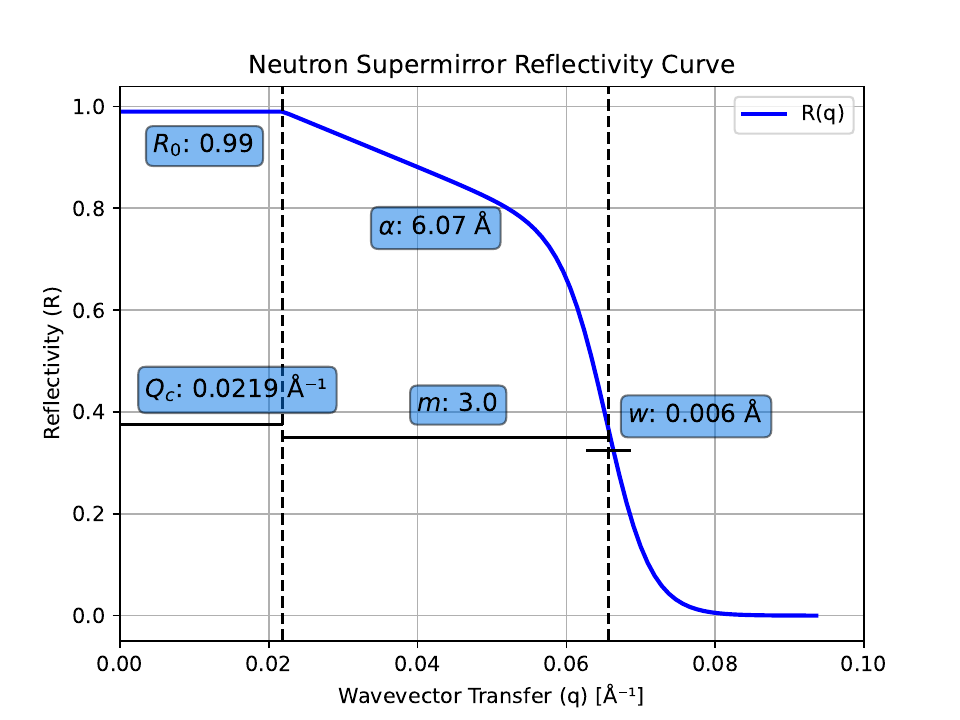}
  \caption{Reflectivity curve $R(q)$ is given by equation \ref{eq:McStas_StdReflecFunc} and shows the probability of reflection given an incident wave vector.}
  \label{fig:refl_plot}
\end{figure}

\noindent{where $Q_c$ represents the cutoff wave-vector, for which neutrons with incident velocity lower than $Q_c$ undergo near perfect reflection; $R_0$ represents the reflectivity for incident wave-vectors $q \leq Q_c$; $\alpha$ describes the slope of the supermirror reflectivity between $Q_c \leq q \leq m \cdot Q_c$; and $m$ shows the supermirror extent in multiples of the $Q_c$ of Nickel \cite{krist_neutron_2008}. For $Q_c \leq q \leq m \cdot Q_c$, neutrons undergo diffraction resulting from stacked multilayers of  Ni/Ti \cite{hayter_discrete_1989}.
The width of the interface region $w$ describes how $R(q)$ decays between the supermirror regime and the regime where incident wave-vector is too great for reflection. An example of reflectivity curve is shown in Figure \ref{fig:refl_plot}
 for a typical supermirror.}

For reflectivity parameter variation and optimization, only first order parameters $R_0$, $\alpha$, and $m$ were varied while $Q_c$ and $w$ were held constant at their original values.
These parameters are specific to each guide component surface specified in the simulation, and precisely dictate the reflectivity performance of that component.

\begin{figure*}
  \centering
  \includegraphics[width=\textwidth]{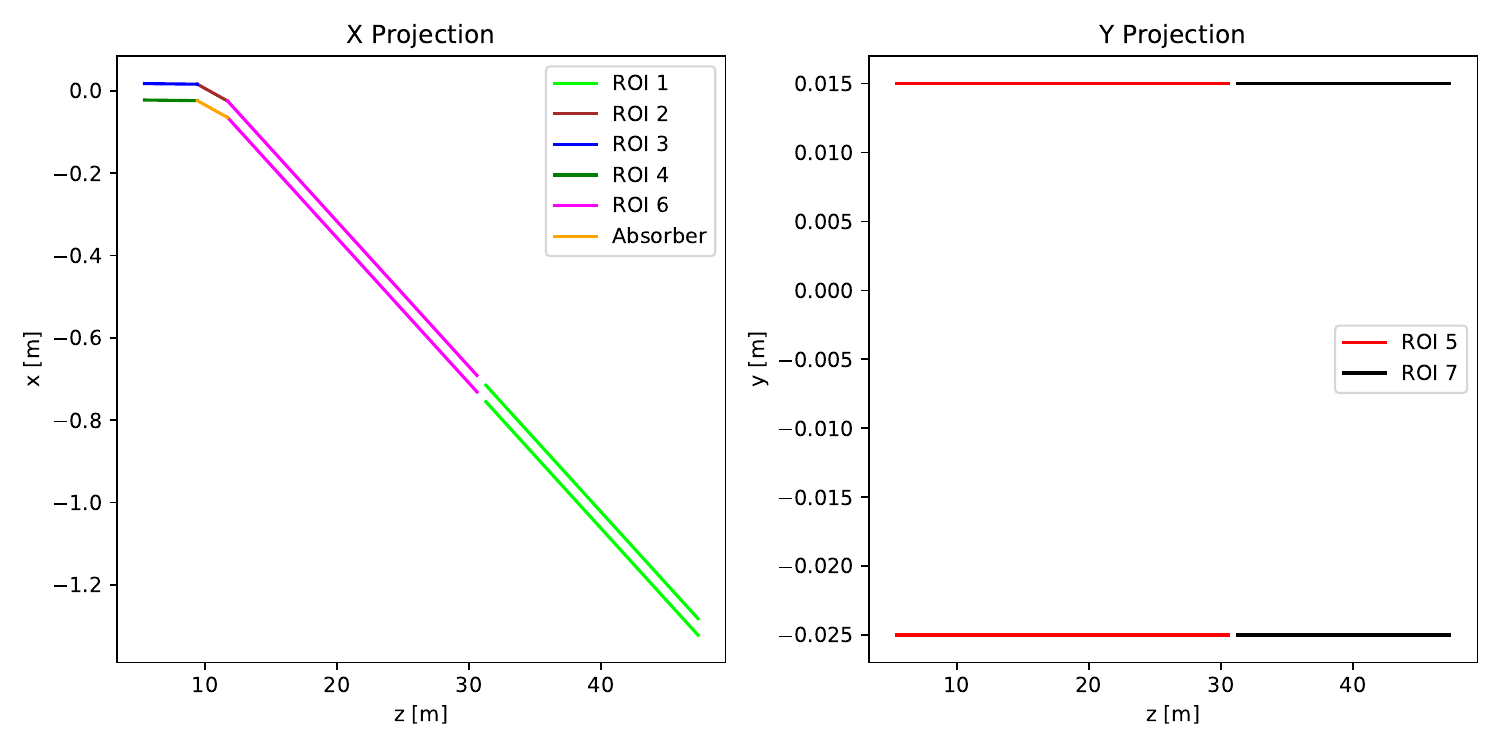}
  \caption{X and Y projections of ROI definitions within the multiparameter optimization.}
  \label{fig:ROI_plots}
\end{figure*}

The CG-2 instrument layout is shown in Figure \ref{fig:HB-4_and_CG-2}.
To consolidate the parameters and help simplify any adjustments that are made to the simulation, regions of interest (ROI) are shown in Figure \ref{fig:ROI_plots}.
These different sections portion out the different optical components of the beamline.
Vertically reflecting surfaces were divided into regions ``upstream'' and ``downstream'' of the Velocity Selector to account for variable exposure along the track, while horizontally reflecting surfaces were divided into 5 different regions to consider the effects of the optical filter directing the pathway to the right.
Region 1 comprises the horizontally reflecting surfaces of the 8 removable collimators between the velocity selector and sample position.
Region 2 is the only horizontally reflecting surface of the optical filter.
It is unique because it is the only ROI whose initial reflectivity parameters were greater than one.
Regions 3 and 4 were the left and right horizontally reflecting surfaces of guide 1, respectively. 
This division was made to account for the possibility that proximity to the source could cause uneven degradation of the guides.
Region 5 comprises the vertically reflecting surfaces of guide 1, the optical filter, and guide 2.
Region 6 contains both horizontally reflecting surfaces of guide 1.
Region 7 comprises the vertically reflecting surfaces of the 8 removable collimators, and is separated from ROI 5 to isolate components downstream of the velocity selector, decoupling optimizations using velocity selector and white beam measurements. Table \ref{tab:originalSimParams} describes the initial reflectivity parameters for each of the different ROIs.  

\begin{table}
  \begin{center}
    \label{tab:originalSimParams}
    \begin{tabular}{c|c|c|c|c|c|c|c}
      \toprule
      \textbf{ROI} & \textbf{1} & \textbf{2} & \textbf{3} & \textbf{4} & \textbf{5} & \textbf{6} & \textbf{7}\\
      \hline
        $R_0$ & 0.99 & 0.99 & 0.99 & 0.99 & 0.99 & 0.99 & 0.99 \\
        $\alpha$ & 6.1 & 6.1 & 6.1 & 6.1 & 6.1 & 6.1 & 6.1 \\ 
        $m$ & 1 & 3 & 1 & 1 & 1 & 1 & 1 \\
      \hline
    \end{tabular}
  \end{center}
  \caption{The original reflectivity parameters for each ROI specified in Figure \ref{fig:ROI_plots}. $R_0$ was bound by [0,1], while $m$ was bounded by [1,3] for ROI 2 and held constant at 1 for all other ROI. The second order parameters where were held constant at $Q_c$ = 0.0219, $w$ = 0.002.}
\end{table}

\subsubsection{Misalignment Parameters}
\begin{figure}
  \centering
  \includegraphics[width=\linewidth]{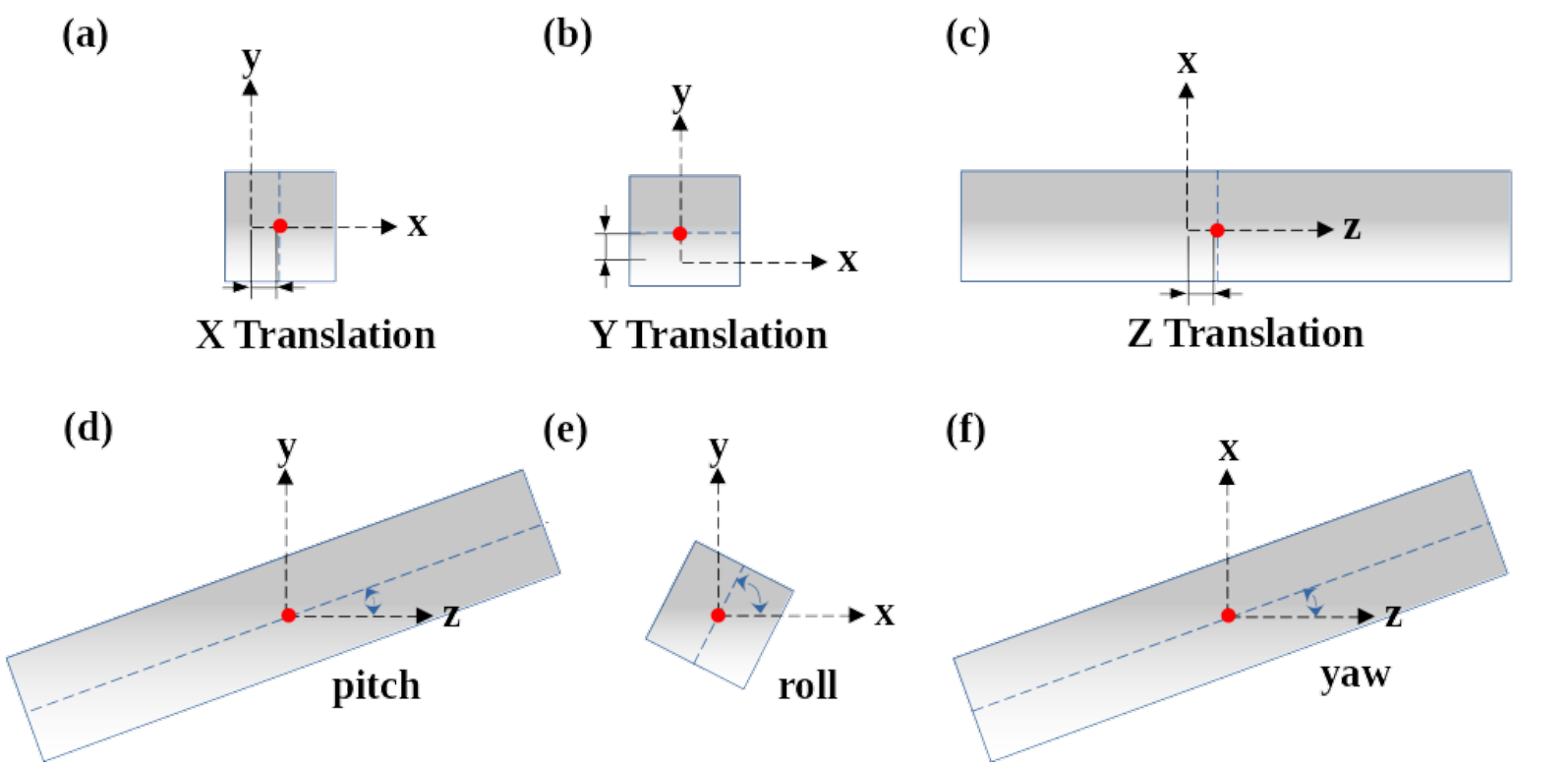}
  \caption{Translational misalignments of guides are shown at top, with (a) X translation horizontal to the beam axis, (b) Y translation vertical to the beam axis, and (c) Z translation being along the beam axis.
  Rotational misalignments of guides are shown at bottom, with (d) pitch misalignments in the vertical plane, (e) roll misalignments around the beam axis, and (f) yaw misalignments in the horizontal plane.}
  \label{fig:misalignment_types}
\end{figure}

Six degrees of freedom were given to each component to model translational and rotational displacement, specified in relation to its original position rather than its position relative to other displaced components. This is shown in Figure \ref{fig:misalignment_types}.
Angular misalignments such as pitch, yaw, and roll were modeled by simply rotating the components through some angle about their center position, and increased angular precision for longer guides was not considered. 

\subsection{Optimization Method}
To minimize the ratio between simulated and measured count rates, the error function for optimization searches was defined as the sum of the squared difference between the ratio of count rates and 1.
\begin{equation}
   \centering
   \text{Err} = \sum_{\text{configs}}\left(\frac{I_{\text{sim}}}{I_{\text{exp}}} - 1\right)^2
   \label{eq:error_function}
\end{equation}

It was necessary to place boundaries on valid parameter ranges to constrain the optimization to realistic degradation possibilities.
If parameters strayed outside of the bounds described in the caption of Table \ref{tab:originalSimParams}, then the amount it strayed by was multiplied by a constant factor of the same order of magnitude as the function described in equation \ref{eq:error_function}, and this penalty was added on top of the error for optimization.
This ensured that the optimized parameters stayed within physically plausible ranges.

\begin{figure}
  \centering
  \includegraphics[width=\linewidth]{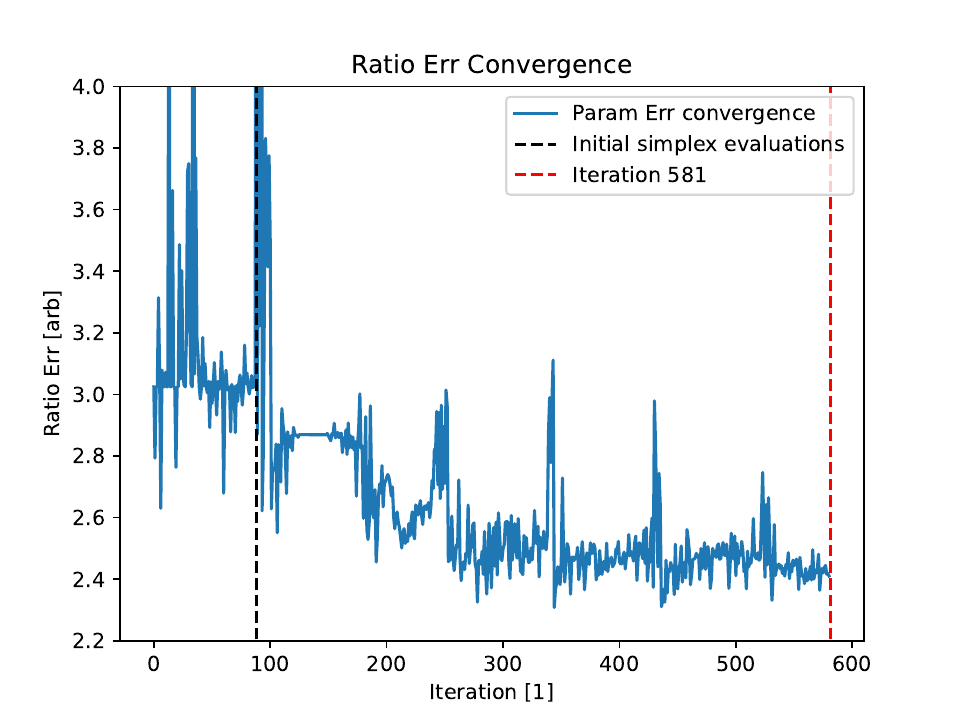}
  \caption{Convergence of optimization error for an example optimization run. Periodicity of ~90 iterations is observed.}
  \label{fig:Optimization_convergence_periodicity}
\end{figure}

Simplex optimization was chosen for this application because it was the simplest method and no derivative information was available \cite{nonlinear_system_identification}. 
Even for the largest parameter set, it only took 4 days with 2-4 minutes per iteration, coming from the need to evaluate each experimental configuration compared for each iteration. 
There did appear to be a periodicity in the convergence, suggesting the whole simplex had to step down to a new level before more progress could be made.
Figure \ref{fig:Optimization_convergence_periodicity} shows this periodicity, with a period that seems to be nearly equal to the number of parameters.


It would be possible to implement more intelligent methods for faster convergence, but for the current size of the parameter set and the configurations per iteration, this method is sufficient.

\subsection{Initial Exploration}
An initial grid search of the reflectivity parameter space was conducted to determine reasonable parameter bounds, which were established by incorporating theoretical limits for each type of degradation. 
Using the McStas implementation of reflectivity curves seen in Equation \ref{eq:McStas_StdReflecFunc}, $R_0$, $\alpha$, and $m$ were varied to model only reflectivity degradation.
$R_0$  was bounded  to [0, 0.99] as 0.99 was the original simulation value.  $\alpha$ has no effect for neutron mirrors with m $\leq$ 1 except for  ROI 2, where $\alpha$ was bounded by [0, 10].
For $m<1$, Equation \ref{eq:McStas_StdReflecFunc} multiplies $Q_c$ by $m$ and sets $m=1$ for further calculation, essentially modeling cutting into $Q_c$.
This is unlikely to impact the specular reflection channel, so only $m$ associated with ROI 2 could vary between [1,3], further limiting the parameter space to investigate.

Misalignment parameter bounds were determined based on rough tolerance estimates along with knowledge of misalignment procedures. 
Rotational misalignment limits were chosen to be 0.2 degrees for pitch, roll, and yaw considering the tolerances found in Table \ref{tab:misalignment_tolerances} and the unlikeliness of such poor alignment.
Translational misalignment limits were chosen to be 0.5 mm for x, y, and z translation considering tolerances found in Table \ref{tab:misalignment_tolerances}.

\subsection{Parameter Optimization Searches}
Several optimization searches were carried out to get estimations for degradation characteristics using available data.
Starting with only white beam data, a simplex optimization of the preferred reflectivity parameters was able to converge within a day and accurately model the observed degradation.

Next, the optimization was expanded to include data from measurements taken with a velocity selector to allow for adjustments to be made to the misalignment parameters.
The final optimization runs each included 75 variable parameters with 48 experimental count rates to compare, usually converging within 1-4 days. 
Although the expanded parameter set reduced the speed of the simplex optimization method, it remained the most attractive choice because it was a simple non-gradient based nonlinear optimization scheme, and our process of evaluating the simulation count rate for different parameter configurations did not provide derivative information. 

\subsection{Parameter Tolerances}
\begin{figure} 
  \centering
  \includegraphics[width=\linewidth]{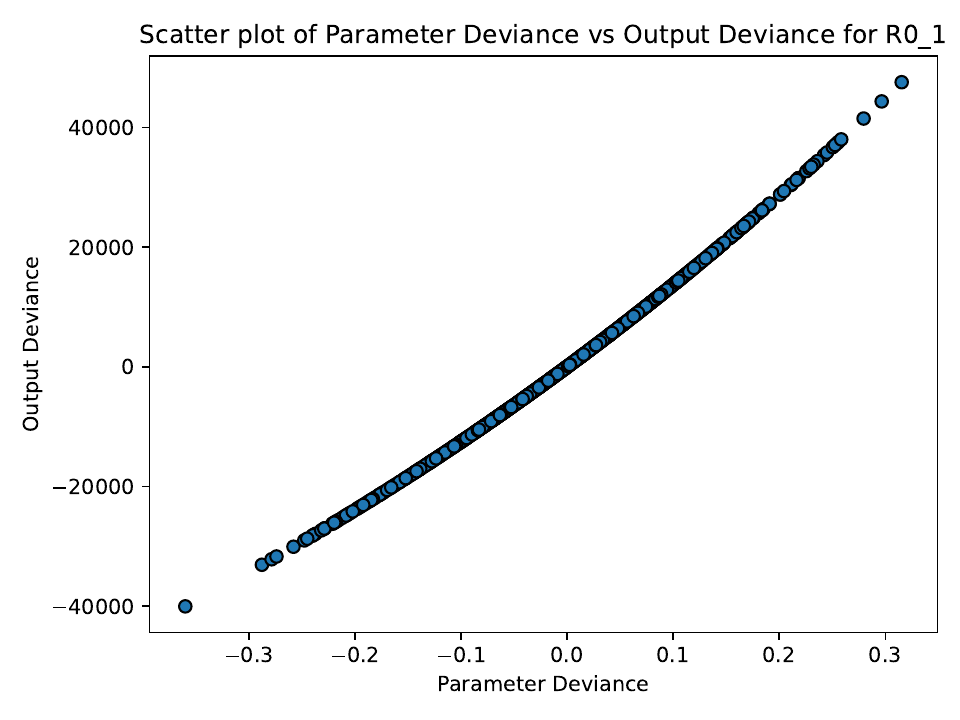}
  \caption{A plot describing the correlation between changes in the maximum reflectivity and the simulation output. When modifying the input $R_0$ to 0.99 plus the ``parameter deviance``, the simulation output changes by the corresponding amount ``output deviance``.
  For 1000 samples, the correlation between input and output deviances is high, $>$99.8\%.}
  \label{fig:parameter-output_correlation}
\end{figure}

The importance of each parameter was determined by analyzing how the output count rates for a white beam with a 20mm diameter aperture and 8 collimators would change given a small change in each parameter around the original parameter set values.
Figure \ref{fig:parameter-output_correlation} shows how the count rate for the above configuration varies given small changes in the original ROI 1 $R_0$ value.
The strong correlation between parameter and count rate deviance justified taking only one sample for each parameter for the final analysis.
This was due to the fact that the magnitude of the slope near the original values can be well approximated with the original output and just one sample point.

To determine the tolerances for the reflectivity parameters, all $R_0$ values, $\alpha$ values, and $m$ values groups were independently adjusted by a range of degradation factors. 
The degradation was performed as a percentage of the optimized parameter values, using global constants ($k_1$, $k_2$, $k_3$) that were applied to each ROI equally. 
Linear fits were performed within regions close to the desired parameter (optimal or original) to establish the parameter tolerances.

\begin{figure}
  \centering
  \includegraphics[width=\linewidth]{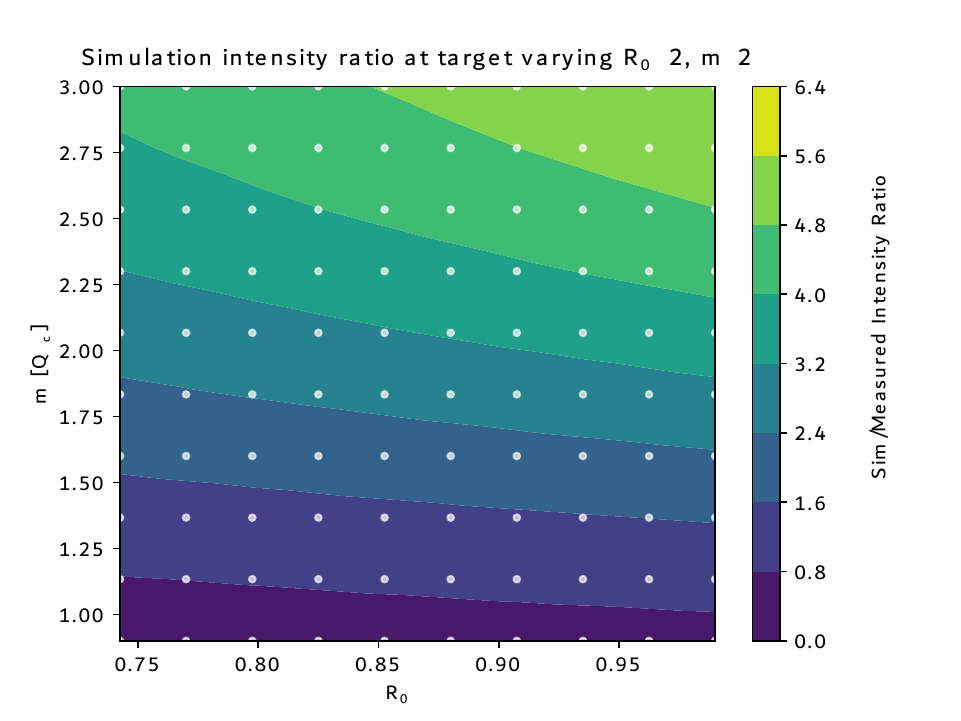}
  \caption{A plot that shows how reflectivity ($R_0$) and critical edge ($m$) parameters impact the simulation output and its comparison to the measured intensities. ROI 2 reflectivity parameters were varied in a grid search using degradation factors ($k_1$, $k_2$, $k_3$), and  simulated instrument count rate at the target was evaluated for white beam configurations. 
  The ratio of simulated count rate to experimental white beam measurements for each tested grid point (gray) was interpolated to find the colored contour map.}
  \label{fig:ROI2C}
\end{figure}

Figure \ref{fig:ROI2C} shows the resulting contour when degradation factors are applied to only the $R_0$ and $m$ values of ROI 2. 
The resulting contour plot shows how reflectivity degradation of the optical filter alone theoretically impacts the performance ratio, with a value of 1 corresponding to the best average match for white beam measurements. 
When parameter tolerances were determined, these factors were likewise applied equally to the individual parameters all regions of interest.

In the region around our original parameters, we considered multiple alignment tolerance values, each representing a range of acceptable variation for misalignment parameters, as inspired by prior work showing the potential impact of mechanical settling of guide components over time \cite{lin_realistic_2023}.
For each tolerance value, we randomly generated sets of 20 parameter samples, drawn from a flat distribution within the defined tolerance limits, [-tol, +tol].
For rotational misalignments, tolerance values considered were 0.020°, 0.075°, and 0.200°. 
For translational misalignments, tolerance values considered were 0.01mm, 0.10mm, 0.50mm, 1.00mm, and 2.00mm.
These randomly sampled parameters were then used to run the simulation 20 times, resulting in a collection of modified spectra. 
To quantify the effect of parameter variations, we averaged the obtained spectra within wavelength bins and subsequently divided the average spectrum by the original, unaltered target spectrum. 
The resulting spectrum ratio plots allowed us to visually and quantitatively analyze the influence of each tolerance on the spectral output. 
To assess the overall impact, we calculated the integral of the spectrum ratio plot, providing a measure of degradation caused by parameter variations. 
Finally, we conducted linear regression fits in regions proximate to the desired parameter values to determine acceptable tolerances for misalignment parameters. 
This rigorous approach not only enables us to understand the sensitivity of our system to parameter variations but also assists in establishing precise tolerance limits, crucial for ensuring the reliability and stability of our system in practical applications.
\section{Results \label{sec:results} }


\subsection{Final Optimization}
The final optimization runs were separated into 2 groups to reduce the parameter set and number of evaluations per iteration. 
The first optimization modified 25 parameters upstream of the velocity selector and compared them against the 36 experimental measurements in the velocity selector spectrum.
With these first 25 parameters fixed, the remaining 50 parameters in the 8 guide sections downstream of the velocity selector were optimized against the 12 datapoints measured at the sample position detector.
The resulting parameters are described in Table \ref{tab:optimizedSimReflParams} and \ref{tab:optimizedSimMisalParams}.

The final optimization estimates that each ROI had an $R_0$ degradation to the mid 80\% range, except ROI 6, which corresponded to horizontally reflecting surfaces in guide 2.
The $\alpha$ and $m$ values for the optical filter reflecting surface showed more realistic degradation than previous optimizations, with the maximum count rate for the velocity selector runs now lining up with experimental measurements.

\begin{table}%
  \begin{center}
    \begin{tabular}{c|c|c|c|c|c|c|c}
      \toprule
      \textbf{ROI} & \textbf{1} & \textbf{2} & \textbf{3} & \textbf{4} & \textbf{5} & \textbf{6} & \textbf{7}\\
      \hline
        $R_0$ & 0.805 & 0.818 & 0.851 & 0.841 & 0.83 & 0.937 & 0.806 \\  
        $\alpha$ & 6.1 & 9.817 & 6.1 & 6.1 & 6.1 & 6.1 & 6.1 \\  
        $m$ & 1 & 2.916 & 1 & 1 & 1 & 1 & 1 \\ 
      \hline
    \end{tabular}
  \end{center}
  \caption{The optimized reflectivity parameters for each ROI specified in Figure \ref{fig:ROI_plots}} \label{tab:optimizedSimReflParams}
\end{table} 

\begin{table*} %
  \begin{center}
    \begin{tabular}{l|r|r|r|r|r|r}
      \toprule
      \textbf{Component} & \textbf{Pitch} & \textbf{Yaw} & \textbf{Roll} & \textbf{X Trans} & \textbf{Y Trans} & \textbf{Z Trans}\\
      \hline
        Guide 1 & -0.890 & 2.300 & -2.590 & 10.840 & 14.090 & -17.080 \\
        Optical filter & 1.360 & 5.830 & 4.880 & 9.770 & 9.600 & 8.900 \\
        Guide 2 & 1.840 & 3.750 & -26.480 & -15.180 & 8.230 & 8.550 \\
        Coll 1 & 20.445 & 3.020 & -4.943 & 6.491 & 0.693 & 3.695 \\
        Coll 2 & 3.377 & 0.711 & -13.344 & -4.332 & 1.445 & -3.630 \\
        Coll 3 & -2.417 & 1.396 & -0.912 & 0.344 & 6.460 & -4.232 \\
        Coll 4 & -7.988 & -1.215 & -45.979 & 7.528 & 6.239 & -3.804 \\
        Coll 5 & -0.052 & -0.074 & -3.126 & -1.601 & 2.084 & -3.349 \\
        Coll 6 & 2.881 & 1.921 & -1.591 & -3.689 & 2.687 & 4.330 \\
        Coll 7 & -0.521 & 2.595 & 3.625 & -0.850 & 4.305 & -2.780 \\
        Coll 8 & 2.450 & -1.530 & 1.102 & 6.951 & 5.927 & 3.101 \\
      \hline
    \end{tabular}
  \end{center}
  \caption{The optimized misalignment parameters for each component in the instrument, shown in Figure \ref{fig:HB-4_and_CG-2}. Angles are expressed in $[deg]*10^{-3}$ and translations are expressed in [$\mu$m]}
  \label{tab:optimizedSimMisalParams}
\end{table*}
Optimized misalignment parameters show a possible misalignment configuration that could explain the observed degradation, but are not necessarily a hard prediction. 

\subsubsection{Final Optimization Outputs}
\begin{figure}%
  \centering
  \includegraphics[width=\linewidth]{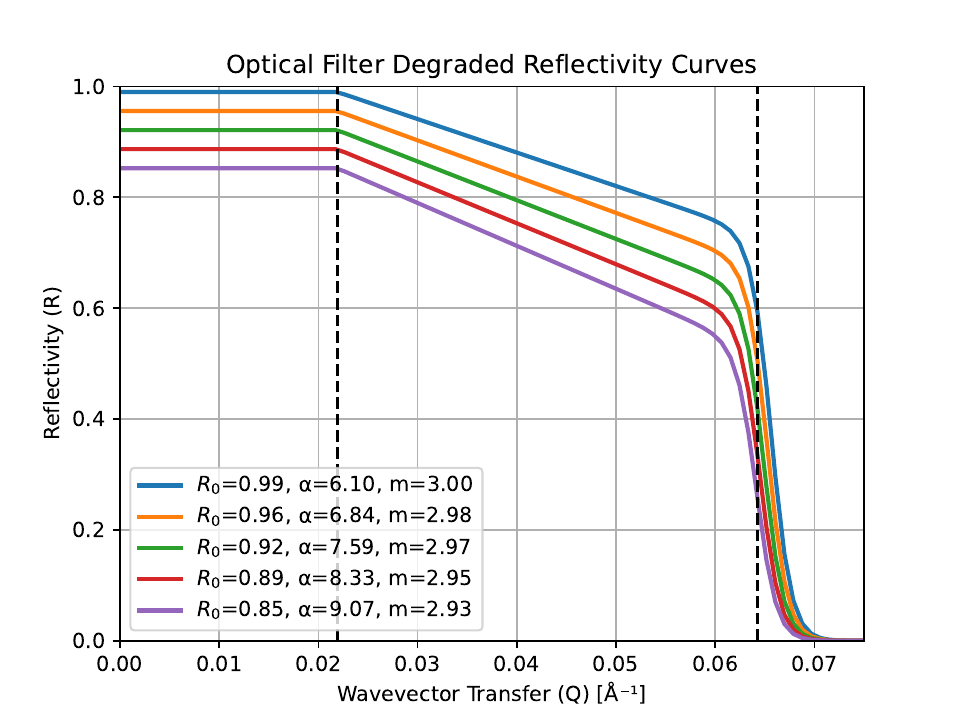}
  \caption{Reflectivity curves using the degradation values from the optimized parameters.}
  \label{fig:Reflectivity_Curve_Degradation}
\end{figure}
Figure \ref{fig:Reflectivity_Curve_Degradation} shows stages of the proposed reflectivity degradation of Table \ref{tab:optimizedSimReflParams}. 
The original curve (blue) loses some of its total area, mostly due to degradation of the $R_0$ parameter until it reaches the final curve (purple). 
Less significantly, $\alpha$ increases slightly, and $m$ decreases slightly.

\begin{figure*}
  \centering
  \includegraphics[width=\linewidth]{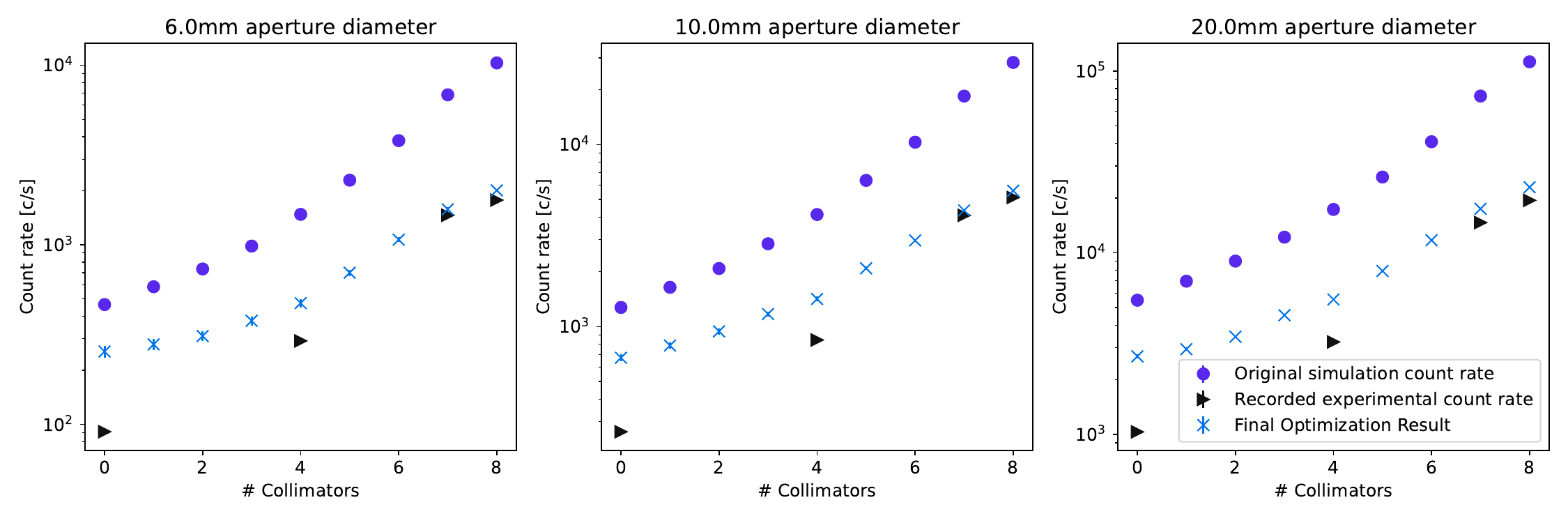}
  \caption{Comparison of white beam measurement data ($\triangleright$) and original simulation count rates ($\circ$) to final optimization simulation countrates ($\times$).}
  \label{fig:White_beam_optimized_simulation_countrates}
\end{figure*}

Original white beam simulated count rates have started to converge to the experimental measurements, with some residual discrepancy among configurations with few collimators around a factor of ~1.3 as can be seen in Figure \ref{fig:White_beam_optimized_simulation_countrates}.  The 'x' show the final optimized count rates which match rather well the o the black triangles.

\begin{figure}
  \centering
  \includegraphics[width=\linewidth]{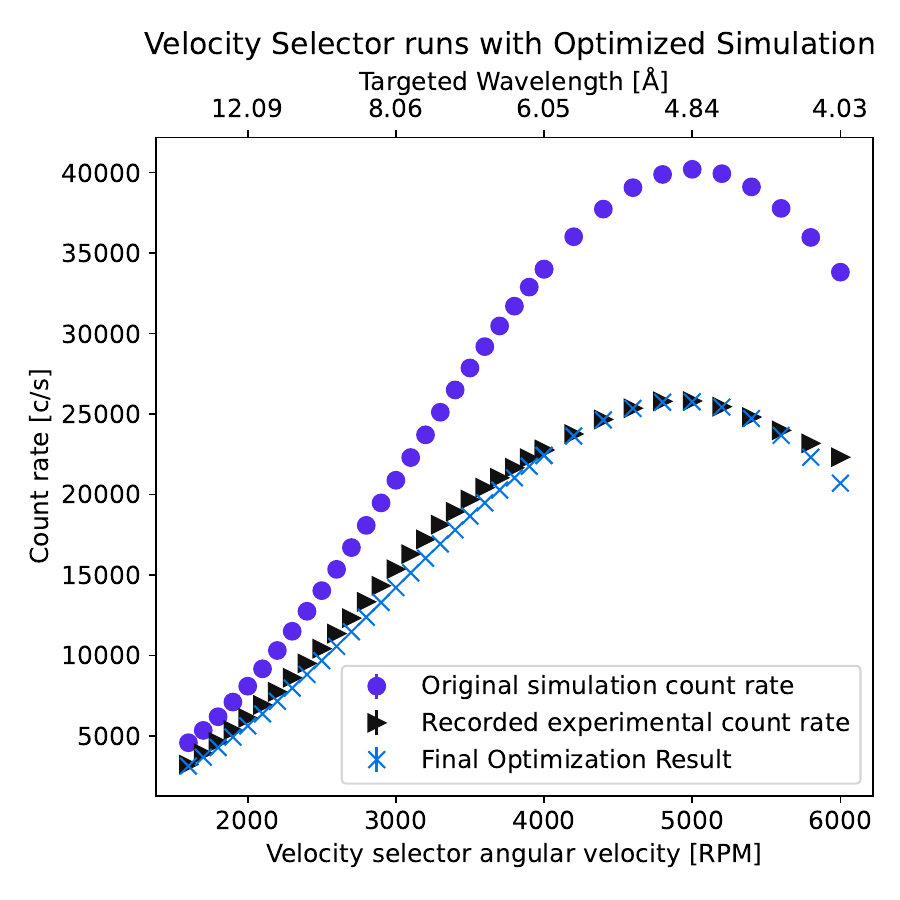}
  \caption{Comparison of velocity selector measurement data ($\triangleright$) and original simulation count rates ($\circ$) to final optimization simulation countrates ($\times$).}
  \label{fig:V_selector_optimized_spectra}
\end{figure}

Optimized simulation count rates at the velocity selector show the same peak value as the experimental measurements, with a slight discrepancy of approximately 1000 cps, or 7\% for Velocity Selector rotation speed between 2500 and 4000 RPM. This is shown in Figure \ref{fig:V_selector_optimized_spectra} where the 'x' data lies almost on top of the measured data.

\subsection{Simulation Variability}
To understand the significance of simulation parameters in influencing the count rate, tests were undertaken to find tolerances near optimal parameters and find which parameters had the largest relative effect on simulation output, as described in section \ref{sec:optimization}.

\subsubsection{Parameter Significance}

\begin{figure}
  \centering
  \includegraphics[width=\linewidth]{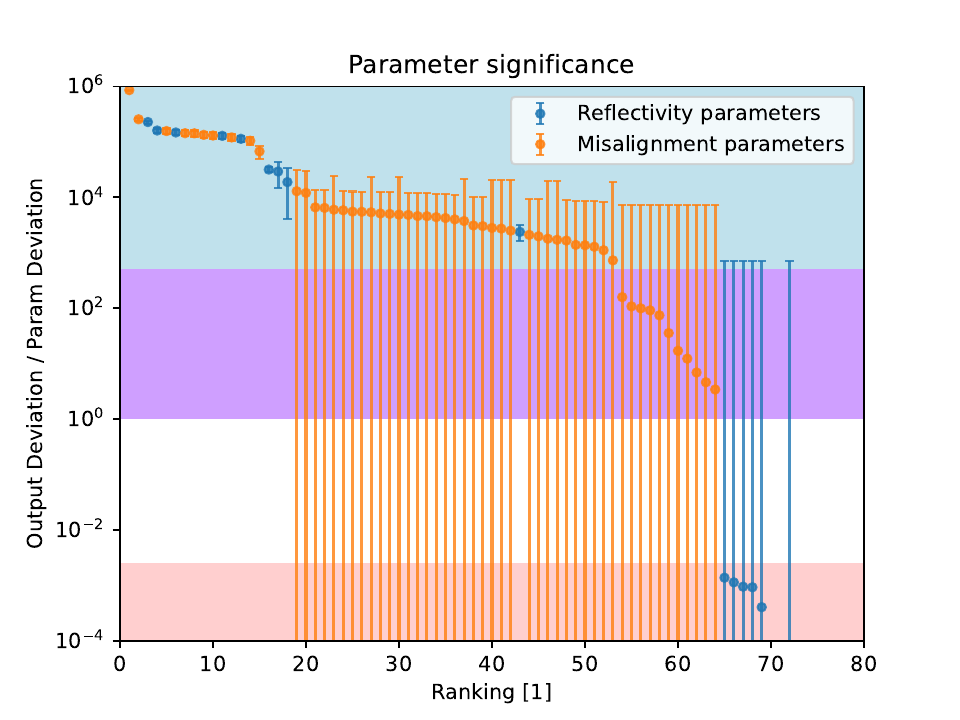}
  \caption{Global ranking of significance of all parameters in descending order. Significance levels were assigned to group parameters, with green showing high significance, purple showing medium significance, and pink showing low significance. Most parameters were highly significant, with only a few reflectivity parameters being practically insignificant.}
  \label{fig:Param_significance_ranking}
\end{figure}

The parameter significance was determined by comparing parameter deviance to target count rate deviance for a single test value, as it was found that parameter deviance and count rate deviance were highly correlated. 
Small parameter deviances were defined as $\delta \approx 0.10 \times Range$ for each parameter type, and simulations were independently run with each parameter deviance test to get a target count rate. 
This count rate was compared to the original count rate with no deviation, and the difference was divided by the parameter deviance to obtain a metric for parameter significance in the region around the optimized parameters.
Figure \ref{fig:Param_significance_ranking} shows the obtained parameter significance metric ranked for all parameters, sorted into 3 regimes. 

The ranking shows that misalignment parameters (orange) dominate the simulation output, with a few reflectivity parameters (blue) in the upper range. $\alpha$ parameters for ROIs with $m=1$ showed negligible significance.

\begin{figure}
  \centering
  \includegraphics[width=\linewidth]{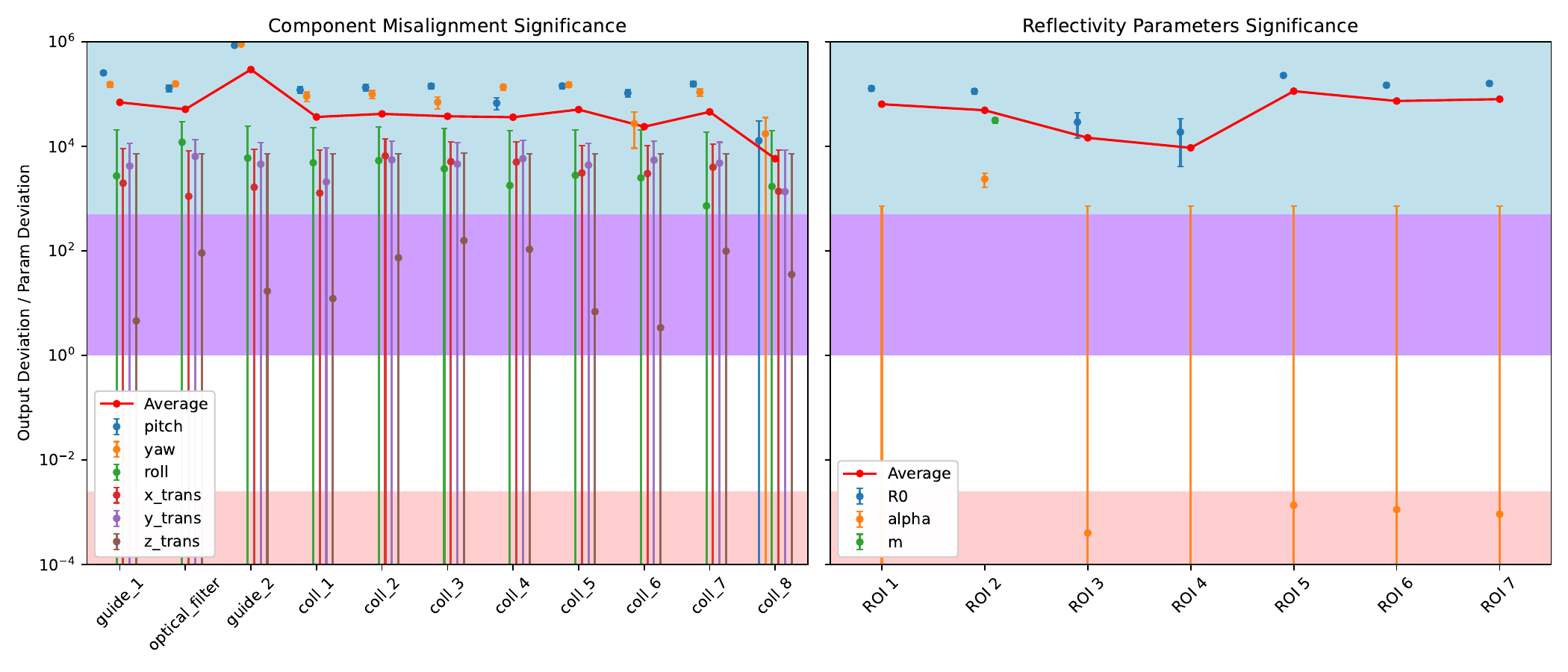}
  \caption{Comparing reflectivity and misalignment parameter significance side by side. Parameter categorization shows that z translation was the least significant parameter excluding $\alpha$ values, but that most other parameters were quite significant.}
  \label{fig:ParamSignificanceComparison}
\end{figure}

Figure \ref{fig:ParamSignificanceComparison} shows the significance of each parameter, sorted into components and levels of significance.
The most significant parameters were found to be angular displacements, especially pitch and yaw, and for Guide 2. 
Almost as significant were the $R_0$ values for most regions of interest, especially ROI 5, corresponding to the vertically reflecting surfaces throughout Guide 1, the Optical Filter, and Guide 2.
X and Y translational misalignments were  the next most significant parameters, approximately the same as angular roll misalignments.
Z translational misalignments were quite low within reasonable bounds.
Excluding ROI 2 for which $\alpha$ significance was on the order of X, Y translational misalignments, $\alpha$ significance was within error of 0. 
This makes sense because all other ROI had $m$ held constant at 1, meaning that $\alpha$ had no impact on the reflectivity curve.

\subsubsection{Parameter Tolerances}
Using degradation factors $k_1$, $k_2$, and $k_3$ as described in section \ref{sec:optimization}, degraded spectra for reflectivity parameters $R_0$, $\alpha$, and $m$ were found in the region around the optimal parameters, in Table \ref{tab:optimizedSimReflParams}. 

\begin{figure}%
  \centering
  \includegraphics[width=\linewidth]{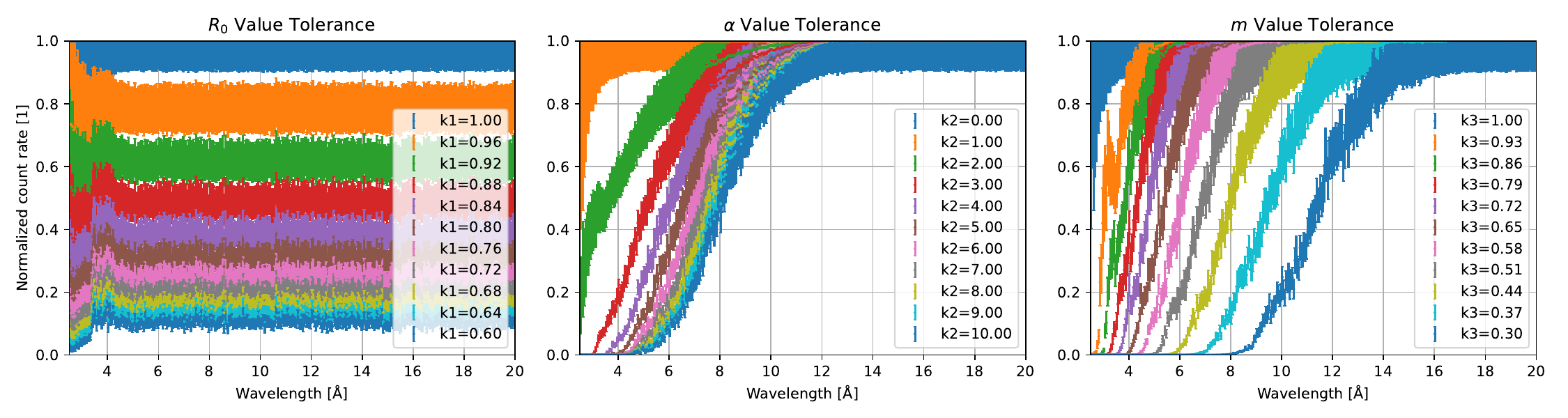}
  \caption{A set of plots comparing the impact of the guide parameters across full spectrum of interest in the simulation. For $R_0$, degradation was relatively constant across the whole wavelength range. For ROI 2 $\alpha$, degradation was not very pronounced until after $\alpha$ had doubled, showing loss in the 2-6 \AA range. ROI 2 $m$ value degradation was more pronounced within the tested range $m$ = [0.87, 2.916], with an increasing drop off in shorter wavelengths.}
  \label{fig:reflectivity_tolerance_spectra}
\end{figure}

As the degradation factor $k_1$ acting on $R_0$ parameter for each guide section was decreased, the count rate across the whole wavelength spectrum decreased without dependence on $\lambda$, meaning that $R_0$ was not was not a wavelength-dependent parameter.
This is shown in Figure \ref{fig:reflectivity_tolerance_spectra}.
Increasing $\alpha$ via $k_2$ causes a proportional decrease in the count rates of shorter-wavelength neutrons, but not as significant a decrease as that of $m$ value degradation.
Decreasing $m$ causes a proportional decrease of shorter-wavelength neutrons with a steeper ``cutoff'' than the decrease observed with $\alpha$ degradation, and in the range of realistic degradation values.

\begin{figure}[b]
  \centering
  \includegraphics[width=\linewidth]{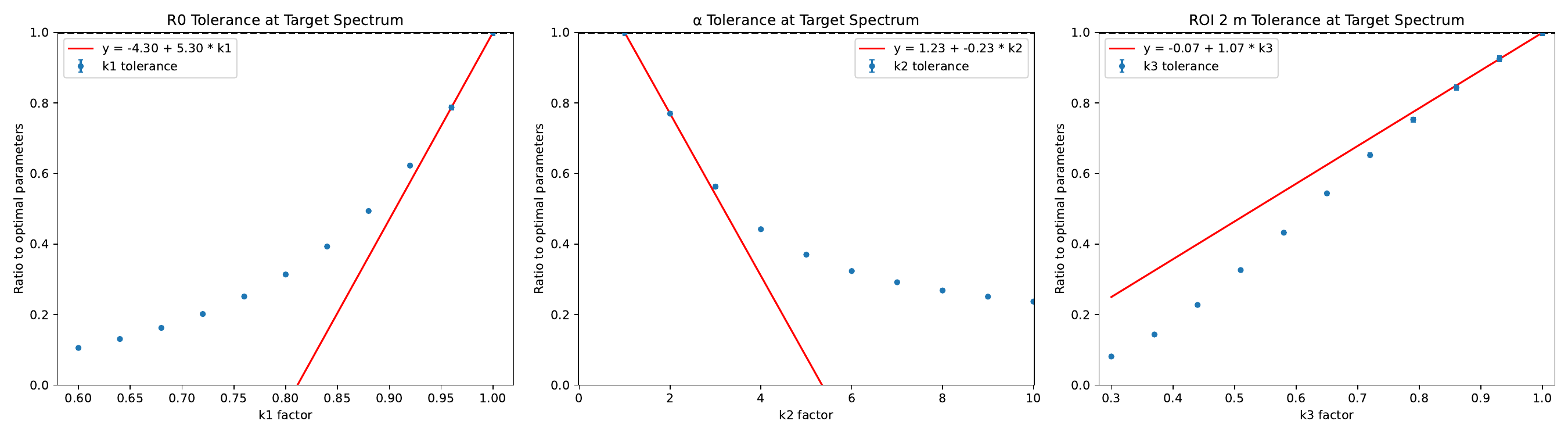}
  \caption{A set of plots showing the impact changing parameters has on the result near the found optimum. The Guide Parameter $R_0$ tolerance was around 53\% loss per 9\% $k_1$ degradation. 
  Guide parameter $\alpha$ tolerance showed degradation of only 23\% when doubling ROI 2 $\alpha$. 
  ROI 2 $m$ tolerance was 11\% loss per 10\% decrease in $k_3$} 
  \label{fig:reflectivity_tolerance_summary}
\end{figure} 

\if {false}
    \begin{figure}%
      \centering
      \includegraphics[width=\linewidth]{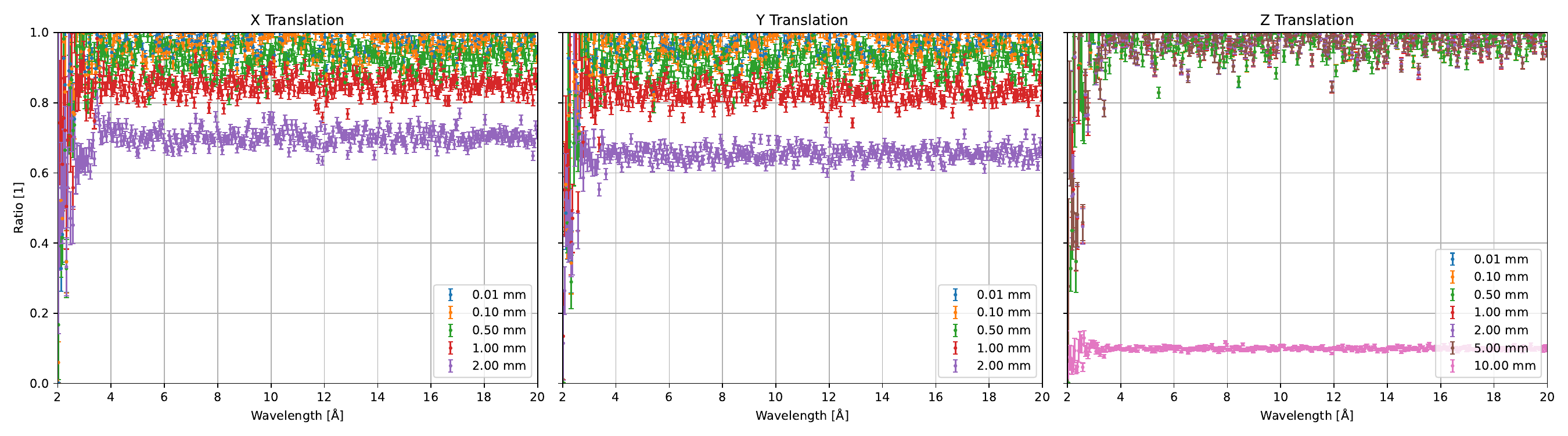}
      \caption{X and Y translation show gradual degradation of instrument performance, while Z translation (along the beam axis) shows a sudden drop off once guides are allowed to form gaps.}
      \label{fig:translational_tolerance_spectra}
    \end{figure}
    
    \begin{figure}%
      \centering
      \includegraphics[width=\linewidth]{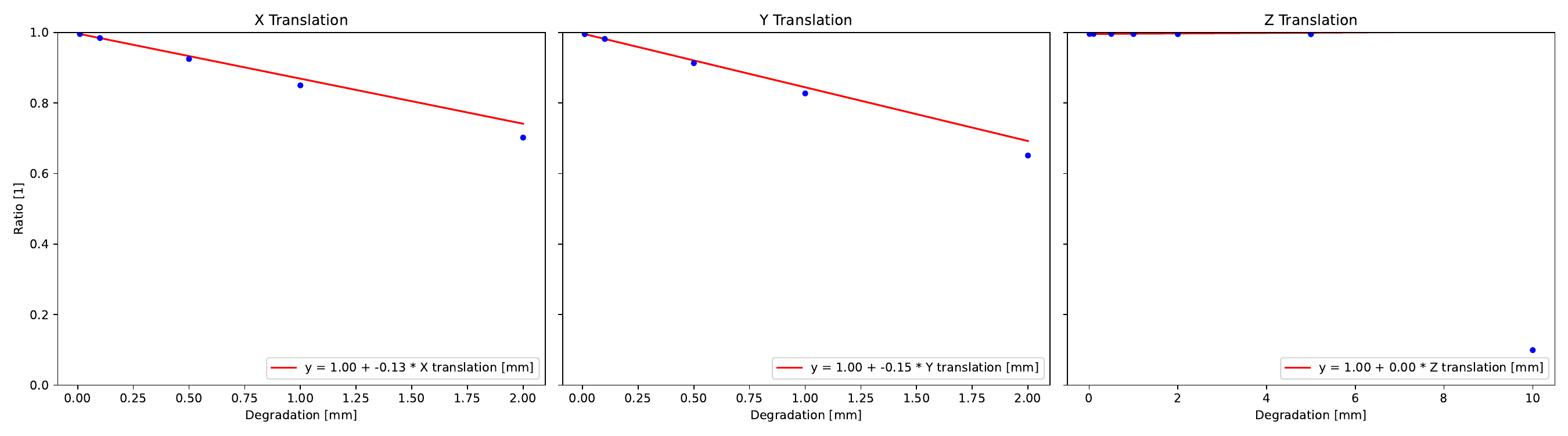}
      \caption{X and Y translational tolerances are moderately tight, with degradation around 14\% per mm tolerance.
      Z translational tolerance is quite loose, allowing up to 5mm tolerance without significant degradation.}
      \label{fig:translational_tolerance_summary}
    \end{figure}
    
    \begin{figure}%
      \centering
      \includegraphics[width=\linewidth]{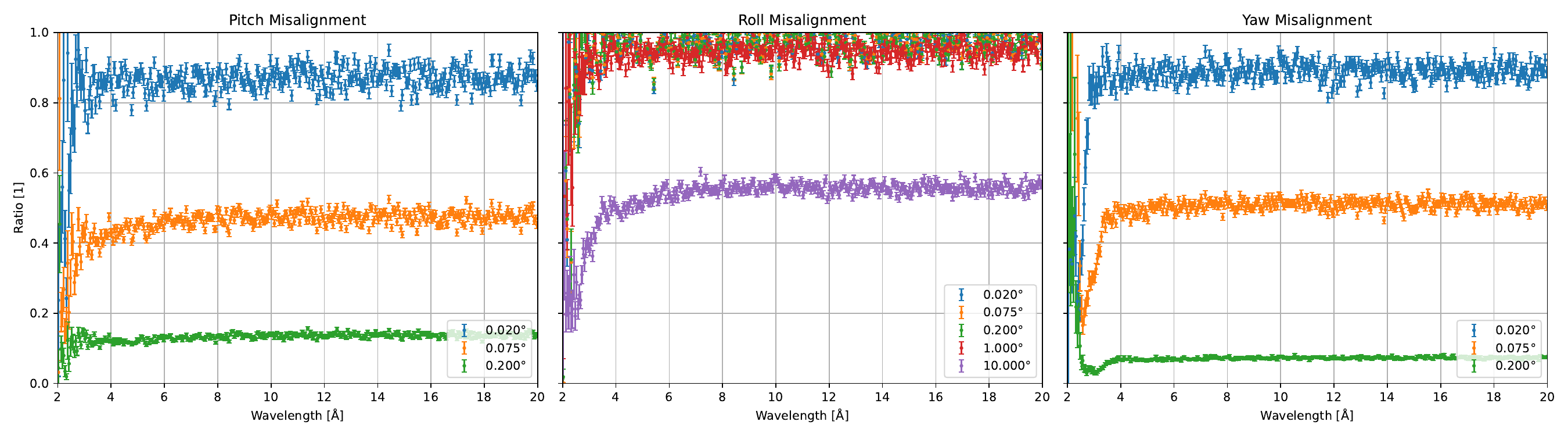}
      \caption{Pitch and Yaw misalignment show similar behavior along the majority of their spectra.
      Roll misalignment did not show significant degradation up to 1° displacement.}
      \label{fig:rotational_tolerance_spectra}
    \end{figure}
    
    \begin{figure}%
      \centering
      \includegraphics[width=\linewidth]{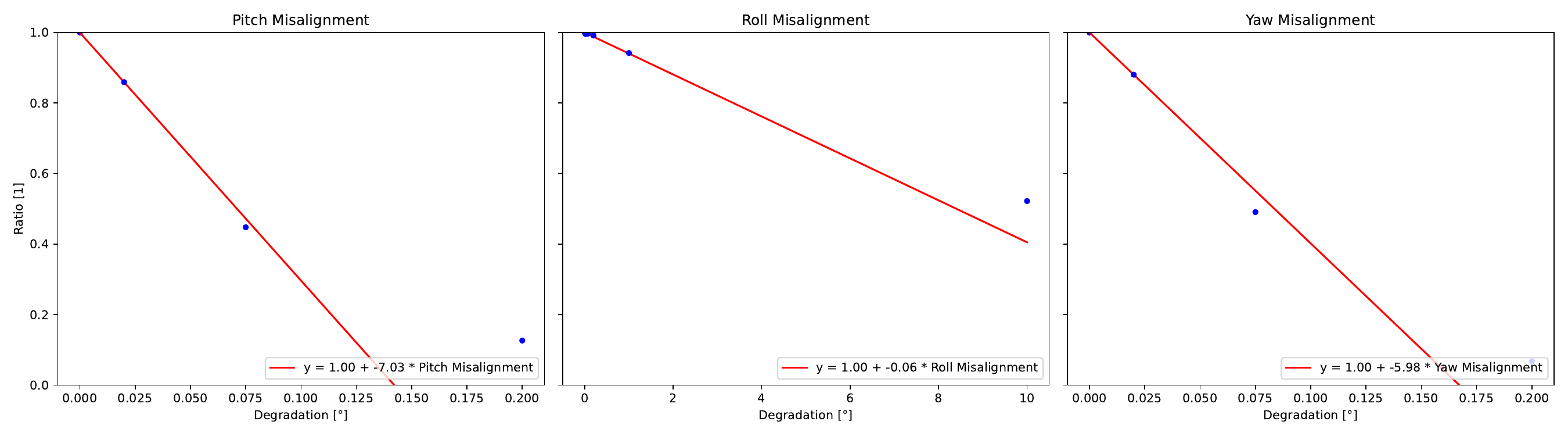}
      \caption{Pitch and Yaw tolerances were quite tight, around 7\% and 6\% per 0.01° pitch and yaw tolerance, respectively. 
      Roll tolerance was much looser, around 6\% per 1.0° roll tolerance.}
      \label{fig:rotational_tolerance_summary}
    \end{figure}
\fi

As described in Section \ref{sec:optimization}, misalignment tolerances were determined via averaging the spectra of 20 white beam simulations at the target position, with components randomly misaligned within given tolerances.

Comparison of the average degraded spectra for these misalignment tolerances showed that unlike reflectivity degradation, losses from misalignments had no significant wavelength dependence.
Furthermore, rotations and translations except for z translation showed close to linear degradation of intensity near the region of the original parameters.
A summary of the found misalignment tolerances can be found in Table \ref{tab:misalignment_tolerances} and is shown in Figure \ref{fig:reflectivity_tolerance_summary}.

\begin{table}
  \begin{center}
    \begin{tabular}{c|c}
      \toprule
      \textbf{Misalignment} & \textbf{Tolerance} \\
      \hline
        Pitch & 7\% per 0.01° \\
        Yaw & 6\% per 0.01° \\
        Roll & 6\% per 1° \\
        X Translation & 14\% per 1mm \\
        Y Translation & 14\% per 1mm \\
        Z Translation & 0\% up to 5mm \\
      \hline
    \end{tabular}
  \end{center}
  \caption{Misalignment tolerances determined from average degradation over 20 random samples within tolerance ranges. Pitch and yaw tolerances were quite tight, with X and Y translation tolerances being less so, and finally Roll and Z translation tolerances being practically ignorable.}
  \label{tab:misalignment_tolerances}
\end{table} 
\section{Discussion \label{sec:discussion} }

Refinement of parameters to match actual measurements to the simulation output was achieved to within 5\% up to the velocity selector data and to within 65\% at the sample position under a white beam configuration.
This result shows that small changes to components that provide the main instrument capability (in this case guide and optical filter components) can have a substantial impact on the instrument simulation.
Given the specialization of those components in particular, quality control of the initial reflectivity and degradation of that reflectivity over time are clearly critical to understanding the potential for an instrument to provide an expected capability throughout its lifetime, as well as ensuring that the instrument simulation reflects the actual performance of the current system.

\subsection{Conclusions}
The proposed misalignment and reflectivity degradation were found to be able to accurately explain the observed discrepancy reasonably well, with the exception of white beam measurements with few collimators.
Likewise, there was difficulty in ensuring wavelength-dependent velocity selector simulations peaked at the same value as measured velocity selector data, but this was found to be resolved with only slight degradation of Optical Filter's $m$ value. 
The discrepancy between the Velocity Selector and White Beam data fits could not be fully reconciled in any of the optimizations or investigations conducted.
This suggests that upstream misalignment was not correctly predicted, or that additional degradation modes should be considered.

Some components were not modeled realistically with degradation parameters in the simulation.
In particular, while the actual Guide 2 is comprised of several 2m long segments, it was modeled as a single component of length 18.9m.
Because of this, simulated misalignment in Guide 2 reflected an unrealistic rotation, and this was likely responsible for the high significance of Guide 2 misalignment parameters.
While it is easy to over complicate incorporating guide misalignments, it would be prudent to at least model Guide 2 with several 2m long segments reflecting the actual instrument geometry.

Determining the relative significance of degradation parameters was helpful in understanding the convergence of optimization, and along with tolerance information it can be used to prioritize future beamline investigations.
It was found that rotational misalignments in long guides were particularly significant (as noted already for Guide 2), followed by $R_0$ reflectivity and then translational misalignments.
Reflectivity parameter tolerances near found optima were quite tight, with approximately 50\% loss per 7\% decrease in $R_0$ value being the tightest tolerance.
The $\alpha$ and $m$ value tolerances for ROI 2 were less tight, with 11\% loss per 10\% decrease in $k_3$ and only 23\% loss observed when doubling $\alpha$.
Misalignment parameter tolerances were found near the original parameters because of the excess of parameters, and they were also quite tight. 
Pitch and yaw were the tightest, nearing 7\% per 0.01° tolerance, although this may be attributed to the extreme length of Guide 2.
X and Y translational tolerances were near 14\% per 1 mm tolerance, while Z translational and Roll rotational tolerances were loose enough to be safely ignored.

\subsection{Future Developments}

While the simulation optimization converged nicely up to the velocity selector, there was substantial deviation in the white beam data taken at the sample position, in particular configurations with fewer collimators.
This suggests the path forward to even better simulation refinement would be another measurement campaign, focused on three measurement types.

First, the same aperture/collimator section intensity scans, but performed with the velocity selector in operation.
These measurements will fill in the dataset, allowing for more precise identification of the discrepancy seen in FIGURE \ref{fig:White_beam_optimized_simulation_countrates}.
Second, measurements of the beam divergence using those same configurations.
The wavelength dependence of the beam divergence distribution will be quite informative with regards to the collimator section reflectivity and alignment of the exit apertures.
Finally, take data into the scattering detector bank with an appropriate sample standard and see if results can be replicated by adding a scattering bank to the simulation that is representative of the actual bank geometry.
This is the ideal end result; a simulation of a scattering instrument whose simulated data output is representative of the actual scattering data seen at the instrument.
That finalized simulation can then be used by potential users to benchmark their sample performance at the instrument, and in some cases, tailor their samples to fit the performance attributes of different instrument configurations that will best meet their experimental goal.

For example, many samples will have prominent scattering features that exist well outside of the $Q$ range of the instrument under some configurations.
If these are pertinent to the structural information of interest by the experimenter, this would directly inform the proper instrument configuration to ensure that relevant scattering data is taken in the correct range.
Furthermore, once a feature is found and if it is located near other dominant structure features, then a coarse resolution configuration could cause a fine feature to be washed out and not be noticed in the data due to the limited resolution of the instrument configuration.
Time would be wasted looking in higher intensity modes for a feature that may be there, but would not reveal itself until a finer resolution configuration is utilized.

Regarding the selected parameter optimization method, the main appeal of simplex optimization for this purpose was its simplicity and lack of derivative evaluation. 
Due to that simplicity, it was easy to customize and implement for this application, with consistent convergence at around 1000-1600 iterations. 
There was a slight excess of parameters which may have contributed to the number of iterations and the periodic behavior described in FIGURE \ref{fig:Optimization_convergence_periodicity}, but overall this was not enough cause to investigate a new optimization method.
Conversely, the main drawback of using simplex optimization for this application is that it is susceptible to getting caught in local minima. 
Since simulation alone is not sufficient evidence to conclude where degradation has occurred, the suggested degradation patterns of any local minima could warrant experimental measurement or direct instrument investigation, and thus finding the exact global minimum is not necessary.

The performance of a wide range of optimization methods were tested in other works, \cite{dijulio_population-based_2016} specifically in the context of neutron optics and shielding design with Monte-Carlo simulations. 
They found that although each optimization had its strengths and weaknesses depending on the particular application, Differential Evolution \cite{storn_differential_1997} had the best average performance. 
Future benchmarking work would likely benefit from the implementation of this method, as it better avoids local minima with little additional overhead.

We propose additional measurement work during the next campaign to include beam divergence, along with testing additional configurations.
The features from these measurements will allow more rich analysis and be better suited to the use of a different optimization method.
More precise identification of the cause of the simulation discrepancy will allow future users to benefit from more precise experiment planning capabilities.



\section{Acknowledgements \label{sec:acknowledgements} }

We would like to thank Lee Robertson for his initial work on the CG2 McStas simulation \cite{cg2_ornl_repo_2023} and Ken Littrell for his help with an internal technical review of this work.
A very special thanks to Thomas Huegle for both his initial work with Lee, as well as a detailed internal technical review of this work.
This research was supported in part by an appointment to the Oak Ridge National Laboratory Undergraduate Research Student Internship, sponsored by the U.S. Department of Energy and administered by the Oak Ridge Institute for Science and Education.
This research used birthright cloud resources of the Compute and Data Environment for Science (CADES) at the Oak Ridge National Laboratory, which is supported by the Office of Science of the U.S. Department of Energy under Contract No. DE-AC05-00OR22725.
This research used resources at the High Flux Isotope Reactor, a DoE Office of Science User Facility operated by the Oak Ridge National Laboratory.

\bibliographystyle{unsrt}
\bibliography{references.bib}

\end{document}